\documentclass[11pt, letterpaper]{amsart}
\usepackage{amssymb}
\usepackage{subcaption}
\usepackage{amsfonts}
\usepackage{amsmath}
\usepackage{amsthm}
\usepackage{fullpage}
\usepackage{setspace}
\usepackage{natbib}
\usepackage{verbatim}
\usepackage{multicol}
\usepackage{multirow,bigdelim}
\usepackage{tikz}
\usetikzlibrary{fit}
\usepackage{graphicx}
\usepackage{xcolor}
\usepackage{multirow, booktabs, dcolumn, color} 
\usepackage{hf-tikz}
\usetikzlibrary{calc}

\tikzset{hl/.style={
    set fill color=red!80!black!20,
    set border color=red!80!black,
  },
}

\usepackage{tikz}
\usetikzlibrary{positioning}

\usepackage{tikz}
\usetikzlibrary{shapes,snakes}
\usetikzlibrary{calc,intersections,through,backgrounds,patterns}

\usepackage{kbordermatrix}

\usepackage{colortbl}
\usepackage[english]{babel}
\usepackage{graphicx}
\usepackage[dvipsnames]{pstricks}
\usepackage{pst-node}
\usepackage{pst-slpe}
\usepackage{lmodern}
\usepackage{multicol}
\usepackage{multirow,bigdelim}
\usepackage{bigints}

\usepackage{graphicx}
\newtheorem{theorem}{Theorem}
\newtheorem{proposition}{Proposition}

\newtheorem{definition}{Definition}

\newtheorem{corollary}{Corollary}
\newtheorem{remark}{Remark}

\begin{document}

\title[]{Coherent Distorted Beliefs\\
}

\author[]{Christopher P. Chambers$^\dag$}
\author[]{Yusufcan Masatlioglu$^\S$}
\author[]{Collin Raymond$^\ast$}
\thanks{$^\dag$ Department of Economics, Georgetown University, ICC 580  37th and O Streets NW, Washington DC 20057. E-mail: \texttt{cc1950@georgetown.edu}.}
\thanks{$^\S$ University of Maryland, 3147E Tydings Hall, 7343 Preinkert Dr.,  College Park, MD 20742. E-mail: \texttt{yusufcan@umd.edu}}
\thanks{$^{\ast }$ Cornell University, \texttt{collinbraymond@gmail.com}. 
\\
We are grateful to Andrew Caplin, Mark Whitmeyer and audiences at BRIC and the 13th Annual Hitotsubashi Decision Theory Workshop.}

\begin{abstract}
Many models of economics assume that individuals distort objective probabilities.  We propose simple coherence conditions on distortions, which ensures that they commute with conditioning  -- in other words it guarantees a form of subjective Bayesianism. Coherence restricts distortions to be power-weighted, where distorted beliefs are proportional to the original beliefs raised to a power and weighted by a state-specific value.  We relate our findings to existing models, including the work of \citealp{grether1980bayes}, inverse-S shaped probability weights, motivated beliefs and multiplier preferences, and the weighted utility of \cite{chew1983generalization}.  Belief-conditional expected utility maximizers with coherent distortions will be dynamically consistent. 
    
\end{abstract}

\maketitle

\section{Introduction}\label{sec:intro}
A key input into decision-making is beliefs about potential states of the world or future payoffs.  A large body of evidence has documented that individuals often display  “belief biases” (see \cite{benabou2016mindful,benjamin2019errors, ortoleva2022alternatives} for recent surveys on distinct aspects of the literature).  Individuals often engage in distortions about the relative likelihood of states of the world as well as the relative likelihood of signals.  These distortions capture many well-known biases --- motivated beliefs (including optimism and overconfidence), conservatism, base-rate neglect, and cognitive uncertainty, to name a few.  

Due to the wide range of evidence pointing towards distorted beliefs, a large number of models have been suggested that can rationalize different forms of distortions. We know of three major branches of the literature.  The first, dating back to seminal work by \cite{edwards1968conservatism},  seeks to explain explain biases in inference and information processing, often in  ball-and-urn style guessing tasks, and draws on both a large literature in psychology and economics (e.g., \cite{grether1980bayes, benjamin2016model, augenblick2021overinference, benjamin2019base} and \cite{ba2022over}).  The second, meant to rationalize a wide variety of violations of expected utility, considers various forms of probability weighting, such as prospect theory (\cite{kahneman1979prospect}), rank dependent utility and cumulative prospect theory (\cite{quiggin1982theory,tversky1992advances}), Choquet expected utility (\cite{schmeidler1989subjective}) and some models of ambiguity aversion such as \cite{maccheroni2006ambiguity}.\footnote{Not all models of ambiguity aversion are representable with a single distorted probability measure, and so are excluded from what we consider models of distorted beliefs.} Third, a more recent literature in economics has modelled individuals who explicitly distort beliefs in an optimistic way, in order to maximize anticipatory utility (subject to some constraint or cost), as in \cite{benabou2002self}, \cite{brunnermeier2005optimal}, \cite{bracha2012affective}, \cite{mayraz2019priors}, and \cite{caplin2019wishful}. Models of distorted beliefs also find direct application in classical economic problems, as in \cite{de2022non}.

Of course, given the range of how distorted beliefs are used in economics, there are a variety of assumptions about the structure of the distortion function which maps the true beliefs about states of the world into the subjective beliefs used for decisions.  Despite the plethora of models, few attempts have been made to understand what kind of natural restrictions we may want to impose on distortion functions.


This paper proposes a class of intuitive restrictions on belief distortion functions, which we call coherency.  Our coherency restrictions capture the intuition that that belief distortions should commute with taking conditionals.  These conditions applies directly to the distortion function itself and thus are applicable to any model that explicitly or implicitly induces probability distortions. 

Our coherency conditions are essentially requirements that agents act as ``subjective Bayesians'' with respect to their distorted beliefs.  This guarantees that individuals consistently distort beliefs, even if they consider counterfactual worlds where they learned some information, or ruled certain states out.  Equivalently, our conditions ensure that our agents are not subject to subjective Dutch books.  The conditions can also be seen as a way of ``robustifying'' models of distorted beliefs with respect to timing: the structure of distortions is the same whether or not the researcher knows that the individual has accessed information prior to distorting or after.  Rather than exploring distorted beliefs in a particular context, our contribution is to explore restrictions on the form of distortion functions, independent of the substantive way in which they are being used (of course, these coherency conditions are more or less compelling depending on the context).  

We show that our conditions imply that distortion functions must take a particular form -- power-weighted. Power-weighted distortion functions correspond to well-known probability distortions found elsewhere in the literature.  They correspond to belief distortions already present in the literature, including Gretherian models where the degree of under inference corresponds to the degree of conservatism, beliefs that emerge from optimal distortions subject to costs proportional to the Kullback-Liebler divergence between the original and distorted beliefs (as in \citealp{mayraz2019priors} and \citealp{caplin2019wishful}, or \cite{strzalecki2011axiomatic}),  and implicit beliefs in the weighted utility  model of \cite{chew1983generalization}.  

Across different sections of the paper, we focus on different kinds of belief distortions.  We start by considering distorted beliefs over states in Section \ref{sec:states}, and then signals in Section \ref{sec:blackwell} separately.  We  then consider situations where the decision-maker engages in joint distortions of the signals and states in Sections \ref{sec:tsd} and \ref{sec:grether}.

In Section \ref{sec:states}, we begin by exploring what we consider the most common form of belief distortion, situations where the decision-maker (DM) distorts probabilities over states of the world, which features in models such as  prospect theory (\cite{kahneman1979prospect}), variational ambiguity aversion (\cite{maccheroni2006ambiguity}) and optimism (\cite{brunnermeier2005optimal}, \cite{bracha2012affective}, \cite{caplin2019wishful}).  A distortion function $\phi$ maps one probability vector (over states) to another.  We focus on mappings that are continuous and positive --- in other words, they map positive probabilities to positive probabilities and zero probabilities to zero probabilities.  We define a distortion function to be coherent if the final belief about a state of the world is the same whether a decision-maker first distorts beliefs, and then conditions on an event $E$, or whether they condition on $E$ and then distort. The following commutative diagram illustrates the different paths: first distortion and then information versus first information and then distortion. In other words, the timing of providing information does not matter for coherent distortion functions. 

\begin{center}
\begin{tikzpicture}
    \node (E) at (0,0) {$\ \ \ p \ \ \ $};
    \node (F) at (4,0) {$\phi(p)$};
    \node (A) at (0,-2)  {$\ p(\cdot|E) \ $};
    \node (B) at (4,-2) {$\ \ q \ \ $};
    \draw[->,double] (E)--(F) node [midway,above] {$\phi$ };
    \draw[->] (E)--(A) node [midway,right] { } 
                node [midway,left] {$E$};
 \draw[->] (F)--(B) node [midway,right] { } 
                node [midway,left] {$E$};
    \draw[->,double] (A)--(B) node [midway,below] { } 
                node [midway,above] {$\phi$};
     \end{tikzpicture}
\end{center}

Our first result shows that coherency requires that the distortion function must be power-weighted: distorted probabilities are proportional to the true probabilities raised to some power, which is then weighted by a state-specific value.  In particular, if $\omega \in \Omega$ is a state, with objective (i.e. undistorted) probability $p(\omega)$ then the distortion function must take the form of 

 $$\phi(p)(\omega)=\frac{\psi(\omega)(p(\omega))^{\alpha}}{\sum_{\omega'\in\Omega}\psi(\omega')(p(\omega'))^{\alpha}}.$$

 Such distortion functions feature raising probabilities to a power, as in models of base-rate neglect (e.g., \citealp{benjamin2019base}) as well as power probability weighting functions used in rank-dependent utility or prospect theory (e.g., \cite{diecidue2009parametric}).  However, they also allow for the probabilities of the states to be higher or lower depending on the identity of the state, as in models of motivated beliefs, like \cite{mayraz2019priors}.  Thus, our notion of coherency is consistent with familiar functional forms from several distinct approach to distorted beliefs.  

Section \ref{sec:blackwell} initiates a study of learning given signal realizations, modeled as Blackwell experiments. Here, we consider a distinct approach considered in the literature to belief distortion --- individuals distort their beliefs about experiment probabilities (which features in models such as \cite{mobius2022managing} and \cite{caplin2019wishful}).  In other words, they implicitly alter the chance of a signal, conditional on a given state of the world, as a way of altering the posterior beliefs conditional on a signal.  We define Blackwell signal coherency as the idea that the updating with respect to a set of signals should commute with respect to distortion.  The results here are largely a reinterpretation of the results in Section~\ref{sec:states}, and we obtain a similar functional form restriction: the distorted signal probabilities (conditional on a state of the world) must be power-weighted.  

Many models of belief distortions suppose that the probabilities of states and signals are distorted simultaneously (e.g., \cite{grether1980bayes}).  The next two sections consider two distinct ways of modelling such simultaneous distortions, and what coherency implies for each approach.  

In Section \ref{sec:grether} we reformulate our problem to consider a generalized class of distortions where the DM can distort, independently, both their prior over states and the Blackwell experiment.  We call such functions Grether updates.  They are widely used in models of probabilistic biases, such as base-rate neglect (as in \citealp{benjamin2019errors} and conservatism (as in \citealp{edwards1968conservatism} and \citealp{benjamin2016model}). We extend our coherency conditions to this environment, where they imply that two distinct operations must be equivalent: (i) distorting the prior, distorting the signal, and then Bayesian updating; and (ii) Bayesian updating with true probabilities, and then distorting the posterior. We show that our approach implies that distortions to states must take on the familiar power-weighted form, while distortions to signals take on a non-normalized power weighted form; and the powers of the two distortions are the same.  The latter occurs because we allow distorted signal distributions to not be probability distributions (as is true in many applications).  If we impose this as a further condition, then no signal distortions are possible, and state distortions can only be weighted (i.e. the power is $=1$).  This implies that an individual is coherent if and only if their attitudes toward distorting base-rates is the same as their attitudes towards distorting information.  


In Section \ref{sec:tsd}, we consider joint distortions inspired by the representation of states and signals as described in \citet{green2022}.\footnote{Another recent use of this structure can be found in \citet{brooks2022}.}
Although isomorphic to the set of priors combined with Blackwell experiments, they allow for distinct concerns about distortions, as now the distortion function acts on the entire matrix at once (as opposed to each row separately, as in the previous subsection).  Our coherency condition is a natural extension of that in Section \ref{sec:states}:  we require our distortion to commute with information from signals, or a subset of signals.  Although this condition is itself relatively weak, we also need to impose an additional condition in order to be comparable to the Grether coherency condition imposed in Section \ref{eq:grether}: the distortion of state probabilities is a function of state probabilities alone, an assumption we call marginality. We show that these coherency and marginality  (along with the standard technical ones) imply that the distortion now must be weighted:  distorted probabilities are proportional to the true probabilities weighted by a state-specific value.  The objective state probabilities can no longer be raised to a power.  Such a representation says that all signal coherent distortions must be ``unseen-information'' Bayesian: it is as if the DM receives a private signal about the state of the world and updates their beliefs over the states of the world using that signal




In Section \ref{sec:apps}, we explore how our results relate to well-known model of preferences.  In the rest of the paper we focus on the properties of belief distortions themselves, without reference to how those beliefs are used to formed preferences.  We now show how different assumptions about the underlying structure of preferences, and how they use the distorted beliefs, allow us to link power-weighted distortion functions to widely used models of non-expected utility.  

First, we extend our model to allow for choices over lotteries, we show the imposing continuity implies that induced choices can be described by the weighted utility model of \cite{chew1983generalization}. We thus link coherent motivated beliefs to non-expected utility phenomena such as the Allais paradox. 

We then relate coherent belief distortions to non-expected utility criterion, and in particular a widely studied class of motivated beliefs which seek to capture the stylized fact that individuals often exhibit optimism and overconfidence.  In these models, individuals distort probabilities to overweight potential good outcomes.  We show that our power-weighted distortions can be seen are the output of an optimization problem solved by individuals with motivated beliefs where they face a cost of distortion that is a generalized version of the Kullback-Liebler divergence between the distorted belief and the true belief.  Weighted distortion functions correspond exactly to the cost being the Kullback-Liebler divergence. We thus link a well-known assumption in the motivated beliefs literature (used by, among others, \citealp{mayraz2019priors} and \citealp{caplin2019wishful}) to our coherency conditions (our approach also directly links up with work in ambiguity on multiplier preferences, as in \cite{strzalecki2011axiomatic} and \cite{hansen2001robust}). We then show that under a belief-conditional expected utility criterion, our notion of coherency implies  dynamically consistent behavior, thus linking our ideas of subjective Bayesiansim to consistent planning. Last, we study the limit beliefs of coherent distortions.  We focus on what happens after repeated application of the probability distortions and provide a characterization of what kind of limit beliefs occur. 


\subsection{Related Literature}\label{sec:related}

Substantively we relate to several literatures that discuss how individuals might subjectively distort (or re-weight) objective probabilities. The first literature focuses on distortions as capturing various probabilistic biases, and dates back to at least \cite{edwards1968conservatism} work on conservatism.  The empirical evidence typically takes the form of ball-and-urn guessing tasks.  There have been a variety of approaches seeking to explain not just conservatism (\cite{benjamin2016model}) but also base rate neglect (\cite{benjamin2019base}) and simultaneous over and under inference (\cite{ba2022over}, \cite{augenblick2021overinference}).  The models often allow for both distortions of states as well as signal probabilities (\cite{grether1980bayes}).  

The second literature grew out of violations of the expected utility hypothesis.  Models such as prospect theory (\cite{kahneman1979prospect}), rank dependent utility and cumulative prospect theory (\cite{quiggin1982theory,tversky1992advances}) were meant to capture well-known violations of the Independence axiom, such as the Allais paradox.  Similarly, models such as Choquet expected utility (\cite{schmeidler1989subjective}) and variational preferences (\cite{maccheroni2006ambiguity}) were meant to accommodate the ambiguity aversion observed in the Ellsberg paradox (there is a much larger literature seeking to explain ambiguity aversion, but many of those models focus on individuals who have a set of beliefs, rather than a single, albeit, distorted belief, which is our focus).  

A third literature has focused on explaining the overly optimistic and motivated beliefs observed in many situations.  These models, such as \cite{benabou2002self}, \cite{brunnermeier2005optimal}, \cite{bracha2012affective}, \cite{mayraz2019priors}, and \cite{caplin2019wishful} typically assume that individuals directly gain utility from their beliefs about future outcomes, and so act to distort those beliefs subject to a constraint or cost.  

Linking notions of power-weighted belief distortions to optimization of beliefs (particularly with Kullback-Liebler divergence costs) has featured in concurrent work by \cite{dominiak2023inertial,strzalecki2024variational,yang2023overprecise,caplin2019wishful}.

Our paper contributes to these literatures by providing a new criterion with which to evaluate belief distortion functions: coherency.  We leverage this to understand which models of distortion satisfy this criterion, and so will satisfy our notion of subjective Bayesianism. We also directly relate power-weighted distortion functions to well-known models of preferences.   

Although less obviously related, the formal structure of our problem is closely tied  to a distinct literature on aggregating beliefs and judgements present not in just economics but also mathematics and statistics.  The equation we use for our main result relies on what is called a Pexider equation (after \citet{pexider}, see \citet{aczel}).  To this end our work is mathematically related to \citet{aczel1983}; see also \citet{aczel1987}.  These papers are interested in aggregating such judgments, but a special case is when there is only one judgment to transform.  Equation~\eqref{eq:pexider} in our proof is of this form, and so the solution of it also follows from the work of \citet{aczel1983}.

Formally, our work contributes to the literature on forming a subjective probability which ``commutes'' with respect to other types of probabilistic information.  An early contribution to this literature is the work of \citet{madansky}, who proposes aggregating subjective beliefs in a way which commutes with respect to the application of likelihood functions.  A characterization of such aggregation procedures was established in full generality in \citet{genest1984characterization,genest}, and features atomic probabilities raised to a power, as in our representation theorem.  The technology behind our results is closely related to the the general framework discussed in \citet{genest1984aggregating}. \citet{west} studies a related problem of simultaneous utility and probability aggregation, and also obtains a power representation for individual probabilities. \citet{chambers2010} investigates a situation where a selection must be made from a convex set of priors, and updating must commute with selection. 

In Section \ref{sec:tsd} we introduce a notion of marginality, which is related to a concept introduced in the aggregation literature by \citet{mcconway1981}; see also \citet{genest1984pooling}.




\section{ Distortions of State Probabilities}\label{sec:states} 



We begin by focusing on a DM who distorts states of the world. Distortions of the probabilities of states feature widely in behavioral economics, ranging from models of probabilistic biases (\cite{benjamin2019errors} to motivated beliefs (\cite{caplin2019wishful}), ambiguity (\cite{maccheroni2006ambiguity}) as well as a large class of non-expected utility models (\cite{quiggin1982theory}).

Formally, let $\Omega$ be a finite set of states of the world, and let $\Delta(\Omega)$ be the set of probability distributions.  A \emph{distorted belief} is a map $\phi:\Delta(\Omega)\rightarrow\Delta(\Omega)$.  A distorted belief is \emph{positive} if $\phi(p)(E)=0$ if and only if $p(E)=0$.  We say that $\phi$ is \emph{continuous} if it is a continuous function.

In this paper, we study a special class of distorted beliefs, which commutes with respect to information. In other words, the timing of receiving information does not affect the final belief distortions--- whether the information is given before or after the distortion will not influence the final posterior. Formally:

\begin{definition}
    A positive distorted belief is \emph{coherent} if for all $E\subseteq \Omega$ for which $p(E)>0$,\[\phi(p(\cdot|E))=\phi(p)(\cdot|E).\] 
\end{definition}

Why is coherency a reasonable condition? We believe there are two key reasons.  First, it implies that the DM is a ``subjective Bayesian.''  In particular, suppose the DM, at the beginning of a period, begins with objective beliefs and then applies the distortion.  If the DM does not satisfy coherency and they consider what they would do if they learned information that ruled certain states out prior to the beginning of the period, compared to the end of the period, they would come to different conclusions. Thus, engaging in this counterfactual reasoning would lead them to understand that they are not fully ``rational,'' in that the timing of information will change their belief.  This cannot happen if they satisfy coherency. In this case, the relative distorted likelihoods of states do not depend on the entire set of states the DM considers possible: the distorted likelihood of states A to B does not depend on whether C is possible.  

We interpret this as a form of subjective Bayesianism --- in particular, it means that coherent DMs are immune to classic ``subjective'' Dutch books.\footnote{Notice that if there are objective probabilities, then any distorted beliefs are subject to some Dutch book arguments.  We focus here on a weaker condition when there are not necessarily objective beliefs.}  For example, consider three states $\{\omega_1, \omega_2, \omega_3\}$, and a decision-maker who is a risk-neutral subjective expected utility maximizer. Let $E=\{\omega_1,\omega_2\}$ and $p$ be the initial belief. Suppose the decision-maker is not coherent, and without loss let $\phi(p(\omega_1|E)) < \phi(p)(\omega_1|E)$ (hence $\phi(p(\omega_2|E)) > \phi(p)(\omega_2|E)$). To see that the timing of information affects risk-taking,  consider a bet where the agent wins $x$ if $\omega_1$ is realized, and loses $\frac{\alpha}{1-\alpha} x$ if $\omega_2$ is realized where $\alpha = ( \phi(p(\omega_1|E)) +\phi(p)(\omega_1|E))/2$.  It is routine to show that DM strictly prefers to take this lottery if DM receives the information after she distorts her initial belief. On the other hand, she rejects the bet if she receives the information before the distortion. Coherency rules out this kind of behavior.  

The second justification is that coherent models are robust to a lack of knowledge on the part of the researcher about the timing of information versus distortion.  Individuals often have access to private unobserved information.  If a researcher attempts to elicit beliefs it is difficult to know whether a decision-maker first distorted their belief, then accessed private information to update it, or rather accessed the private information, then distorted their beliefs.  So long as the model of distorted beliefs satisfies coherency, the researcher can be agnostic about which of those two (unobserved) processes took place. Without coherency, as our betting example demonstrated, it is very difficult to predict the actual behavior.

\begin{figure}[htbp]
\centering

\scalebox{0.35}[0.35]{\includegraphics{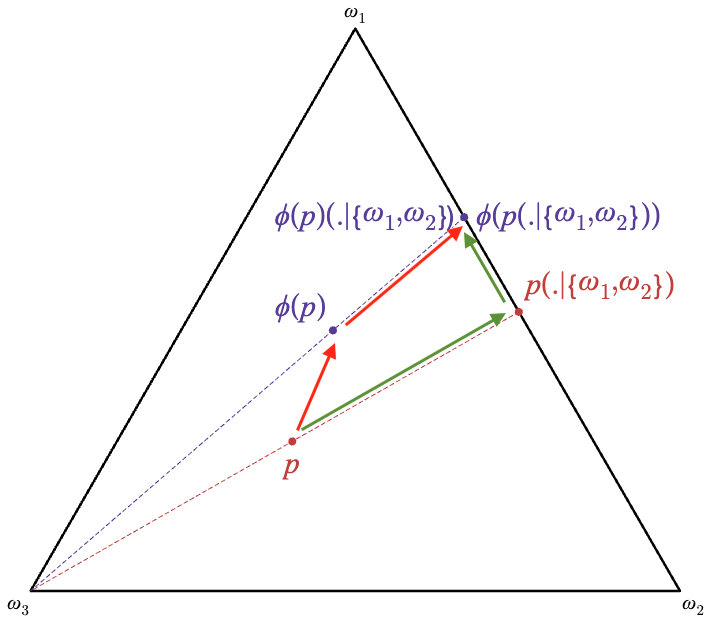}} 
 \caption{\footnotesize{Visualisation of Coherency:  In the absence of any information, $\phi(p)$ depicts the distorted belief of the initial belief $p$. Given the information that the state lies in $\{\omega_1,\omega_2\}$, $\phi(p)(.|\{\omega_1,\omega_2\})$ represents the updated belief of the distorted prior, which is equal to the distorted belief of updated prior, $\phi(p(\cdot|\{\omega_1,\omega_2\}))$. The former is indicated by the red arrows, and the latter is highlighted by green arrows.}}
 \label{fig:distortionfunction}
\end{figure}

\begin{figure}[htbp]
\centering

\scalebox{0.25}[0.25]{\includegraphics{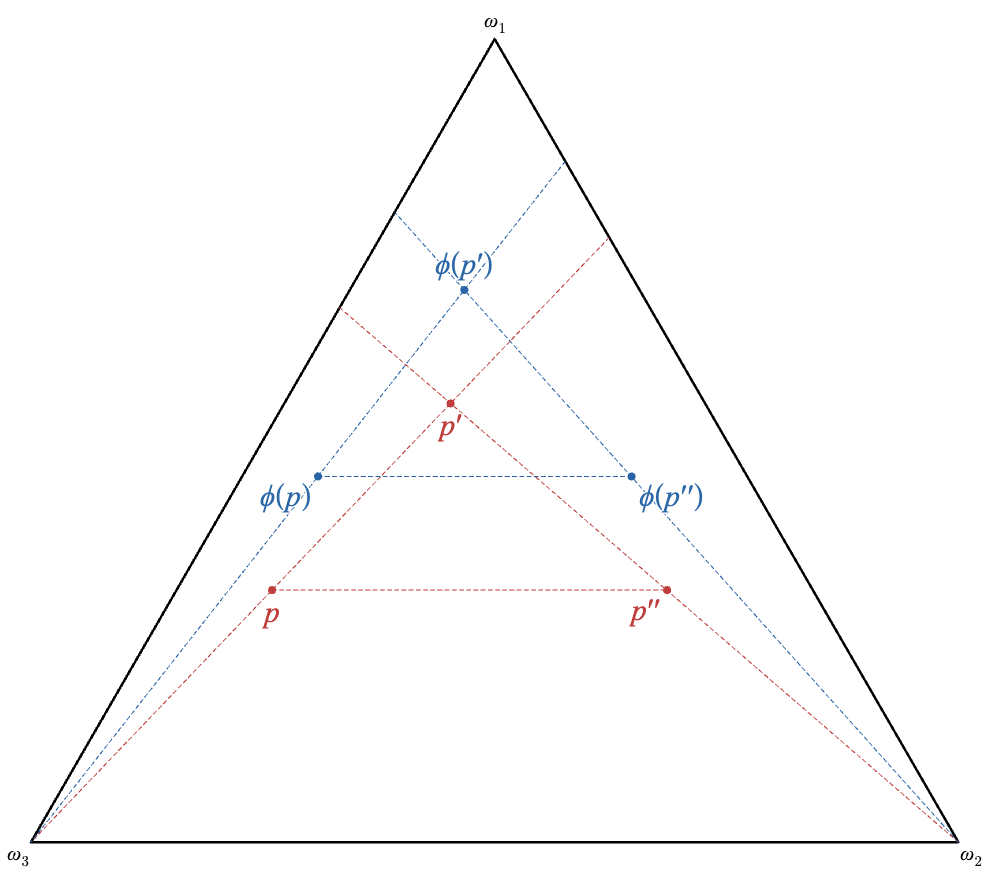}} 
 \caption{\footnotesize{Visualisation of $\Pi$-marginality: Here $\Pi=\{\{\omega_1\},\{\omega_2, \omega_3\}\}$ and  $p(\{\omega_2, \omega_3\})=p''(\{\omega_2, \omega_3\})$, which are on the same horizontal line. Hence, $\Pi$-marginality imposed that $\phi(p)(\{\omega_2, \omega_3\})$ and $\phi(p'')(\{\omega_2, \omega_3\})$ are equal, which implies that they must be on the same horizontal line.}  }
 \label{fig:measurable}
\end{figure}

Figure \ref{fig:distortionfunction} provides a visualisation of a distortion function satisfying coherency.  Suppose there are three states of the world, and there is an initial belief over the states $p$. The distorted belief without information is depicted by $\phi(p)$ inside the simplex.  Now assume that the agent learns that $\omega_3$ does not occur, which is equivalent to event $E=\{\omega_1,\omega_2\}$. There are two ways to approach this new information. In the first, the agent updates the initial belief, that is, $p(\cdot| E)$, which is a point on the $\omega_1-\omega_2$ edge. Then the agent distorts this updated initial belief, which is indicated by $\phi(p(\cdot|\{\omega_1,\omega_2\}))$. In the second, the agent can update the initial distorted belief by using information $E$, that is, $\phi(p)(\cdot|\{\omega_1,\omega_2\})$. Note that $\phi(p)(\cdot|\{\omega_1,\omega_2\})$ is equal to $\phi(p(\cdot|\{\omega_1,\omega_2\}))$, hence it is coherent. Note that the agent distorts her belief towards state $\omega_1$ in both cases. Another interesting implication of coherency can be seen in this figure. Take any initial belief $q$ on the same array as in $p$. Then coherency implies that $\phi(p)$  and $\phi(q)$ must be on the same projection line. Hence, any $\phi$ satisfying coherence maps any projection line to some other projection line (see Figure \ref{fig:measurable}).

To characterize coherent distortions, we next consider a particular class of distorted beliefs called {\it{power weighted}} distorted beliefs:$$\phi^{PW}(p)(\omega)=\frac{\psi(\omega)(p(\omega))^{\alpha}}{\sum_{\omega'\in\Omega}\psi(\omega')(p(\omega'))^{\alpha}}$$ where  $\psi(\omega)>0$ for all $\omega$, and $\alpha > 0$.  $\psi(w)$ is the state-dependent weighting value, and $\alpha$ represents the relative importance assigned to the objective probability.  We introduce a piece of simplifying notation.  For a set $B$ and functions $f,g:B\rightarrow \mathbb{R}$, we use the notation $f(x)\propto_B g(x)$ to mean that there exists some $\lambda > 0$ for which $f = \lambda g$.  When the set $B$ is obvious, we drop the subscript.  As an example, for a power weighted distorted belief with parameters $\alpha$ and $\psi$, $\phi^{PW}(p)(\omega)\propto_{\Omega} \psi(\omega)p(\omega)^{\alpha}$.

We turn to interpreting this functional form.  Suppose $\alpha=1$.  Then the distorted likelihoods of states are weighted towards those states with a high $\psi(\omega)$.  In contrast, if we suppose $\psi$ is constant, then if $\alpha>1$, distorted probabilities are weighted towards the most objectively likely outcomes, while an $\alpha<1$ leads to likelihood ratios becoming closer to 1.  More generally, if $\alpha$ approaches zero, the distorted belief is heavily weighted toward the state-dependent weights $\psi$. As $\alpha$ increases, the distorted belief is determined by the most probable state according to objective belief. In particular, as $\alpha \rightarrow 0$ then the distorted probabilities are determined entirely by $\psi$, and are independent of the objective probabilities.  As $\alpha \rightarrow \infty$ then the distorted beliefs place all weight on the objectively most likely states.

$\phi^{PW}$ possesses desirable properties. First of all, $\phi^{PW}$ is positive as long as $\psi$ is positive and $\alpha>0$. In addition, $\phi^{PW}$ is also continuous. Finally, $\phi^{PW}$ is coherent. To see this, let $p\in\Delta(\Omega)$ and let $E\subseteq \Omega$ for which  $p(E)>0$.  Then $$\phi^{PW}(p(\omega|E))=\frac{\psi(\omega)\left(\frac{p(\omega)}{p(E)}\right)^{\alpha}}{\sum_{\omega'\in E}\psi(\omega')\left(\frac{p(\omega')}{p(E)}\right)^{\alpha}}=\frac{\psi(\omega)(p(\omega))^{\alpha}}{\sum_{\omega'\in E}\psi(\omega')(p(\omega'))^{\alpha}}=\phi^{PW}(p)(\omega|E)$$

The next result establishes that if a distorted belief satisfies these three conditions, it must be written as a power-weighted distorted belief.   This result provides a full characterization of coherent distorted beliefs.\footnote{Proofs are elevated to Appendix \ref{app:proofs}.}  Closely related results appear in \citet{aczel1983} and \citet{genest1984aggregating}, where the coherence property is formulated slightly differently, and where these papers refer implicitly to the full support case.

\begin{theorem}\label{thm:motivated}Let $|\Omega|\geq 3$.  A distorted belief $\phi$ is positive, coherent, and continuous if and only if $\phi$ has a power-weighted distorted belief representation.\end{theorem}

Theorem \ref{thm:motivated} shows that any power-weighted distorted belief must satisfy coherency. More importantly, any distorted belief satisfying coherency (along with positivity and continuity) must have a power-weighted representation. However, Theorem \ref{thm:motivated} does not speak about the parameters of the model. We next discuss how to reveal the state-dependent weights and the power parameters. 

It turns out that uniform belief plays a key role in revealing weights. Let $p_u$ be the uniform belief where $p_u(\omega_i)=p_u(\omega_j)$ for all $i,j$. Then we have    $$\frac{\phi(p_u)(w_i)}{\phi(p_u)(w_j)} =\frac{\psi(\omega_i)}{\psi(w_j)}  $$ which uniquely reveals $\psi$ up to scalar multiplication. Note that if $\phi(p_u)=p_u$, $\psi$ must be also uniform. In that case, $\alpha=1$ if and only if $\phi(p)=p$ for all $p$. If $\phi(p_u)$ is not equal to $p_u$, then there are at least  two states $\omega_i$ and $\omega_j$ where $\psi(\omega_i) \neq  \psi(\omega_j)$. Then  $\alpha$ is determined by the following formula:$$\alpha = \frac{ \ln \phi^2(p_u)(\omega_i) -  \ln  \phi^2(p_u)(\omega_j)}{\ln \phi(p_u)(\omega_i) -  \ln \phi(p_u)(\omega_j)}   -1$$

An interesting special case of power-weighted distorted belief is when $\alpha$ is equal to $1$. Given our identification of $\alpha$, we have the following corollary. 




\begin{corollary}\label{cor:alpha=1} Let $\phi$ have a power-weighted distorted belief representation.  Then $\alpha = 1$  if and only if $$\frac{\phi(p)(\omega_i)}{\phi(p)(\omega_j)}\frac{\phi(q)(\omega_j)}{\phi(q)(\omega_i)}= \frac{p(\omega_i)}{p(\omega_j)}\frac{q(\omega_j)}{q(\omega_i)}$$ for any $p,q$ and $\omega_i,\omega_j$ for which $\min\{p(\omega_j),q(\omega_i)\}>0$.\end{corollary}

 Consider an agent who perceives the world as $\Omega=\{E,\{\omega\}_{\omega\notin E}\}$. This agent does not distinguish between states in $E$. This could be because the agent cannot perceive the difference between states in $E$. Alternatively, it could be that the agent does not believe any further specification of the uncertainty about states in $E$ is necessary, perhaps because these further contingencies are not contractible.  Still, in case any of them potentially become contractible, she wants to ensure no contradictions would arise upon distorting her belief.  Such an individual will want to ensure that how she allocates probabilities across the states in $E$ does not affect her distorted probability of $E$. That is, if $p(E)=p'(E)$, then $\phi(p)(E)=\phi(p')(E)$.  

Formally, let $\Pi$ be a partition of $\Omega$.  Say that $\Pi$ is \emph{nontrivial} if there exists $E^*\in\Pi$ for which $1<|E^*|<|\Omega|$.  Say that $\phi$ is \emph{$\Pi$-marginal} if for any $p,p'\in\Delta(\Omega)$, if for all $E\in\Pi$, $p(E)=p'(E)$, then for all $E\in\Pi$, $\phi(p)(E)=\phi(p')(E)$. $\Pi$-marginality is the condition that the distorted probability of an event in $\Pi$ depends only on the probability of that event, and not how the probability is allocated across the states in that event. Figure \ref{fig:measurable} provides a visualization for $\Pi$-marginality. We assume that $\Pi$ only distinguishes state 1, hence $\{\{\omega_1\},\{\omega_2, \omega_3\}\}$. Then if we have two priors $p$ and $p''$ assigning the same probability $E=\{\omega_2, \omega_3\}$   ($p(E)=p''(E)$). This implies that $p$ and $p''$ are on the same horizontal line. Hence, $\Pi$-marginality imposes that the corresponding distorted beliefs are also on the same horizontal line, $\phi(p)(E)=\phi(p'')(E)$.

\begin{corollary}\label{cor:motivated}Let $\Pi$ be a nontrivial partition of $\Omega$.  Then $\phi$ satisfies positivity, continuity, coherency and  $\Pi$-marginality if and only if $\alpha = 1$ and the map $\psi$ is $\Pi$-measurable.\end{corollary}

 Thus, $\Pi$-marginality forces coherent belief distortions to be of the weighted form. To see this, we first establish that $\Pi$-marginality implies that $\psi$ is $\Pi$-measurable.  The argument roughly fixes $\omega,\omega'\in E\in \Pi$ and $\omega^* \notin E$.  By applying $\phi$ to each of $(1/2)\delta_{\omega}+(1/2)\delta_{\omega^*}$ and $(1/2)\delta_{\omega'}+(1/2)\delta_{\omega^*}$ (each having the same marginal distribution over $\Pi$) the conclusion immediately follows by simple algebra.\footnote{Here, $\delta_{\omega}$ refers to the point mass on $\omega$.}  That $\alpha=1$ follows from a similar trick after this derivation, by considering $(1/4)\delta_{\omega}+(1/4)\delta_{\omega'}+(1/2)\delta_{\omega^*}$ and $(1/2)\delta_{\omega}+(1/2)\delta_{\omega^*}$, observing that these have the same marginal over $\Pi$, and again using simple algebra.  
 

Our first two results put tight conditions on the structure of distortions that are consistent with coherency. Power-weighted distortion functions are conceptually distinct from some other well-known types of probability distortions. Unlike prospect theory distortions, but like models of rank-dependent utility and cumulative prospect theory (and many other models of belief distortions), our distortion operator maps a distribution to a distribution.  Unlike rank-dependent utility, distortions do not depend on the rank of the state.\footnote{See Section \ref{sec:apps} for applying our model to lottery choice, and a discussion of the related continuity concerns.}  


That said, despite the tight functional form restrictions coherency imposes, the class of distortion functions that are allowed correspond to intuitive forms of belief distortions.  For example, when $\alpha<1$, we obtain an inverse-S shaped distortion function, reminiscent of what is often used in models of prospect theory, cumulative prospect theory, and rank-dependent utility (\cite{kahneman1979prospect}; \cite{tversky1992advances};\cite{quiggin1982theory}).  In the world of ambiguity, when $\alpha=1$, as we will discuss in Section \ref{sec:apps}, our approach can be consistent with the induced beliefs found in multiplier preferences as in \cite{strzalecki2011axiomatic,hansen2001robust}; as well as models of motivated beliefs as in \cite{caplin2019wishful,mayraz2019priors}. When $\psi$ is constant, and $\alpha<1$ models of base-rate neglect (\cite{benjamin2019base}) are consistent with our model, as are notions of cognitive uncertainty (\cite{enke2023cognitive}) where the DM uses a uniform distribution as the default.

\begin{figure}[htbp]
\centering

\scalebox{0.2}[0.2]{\includegraphics{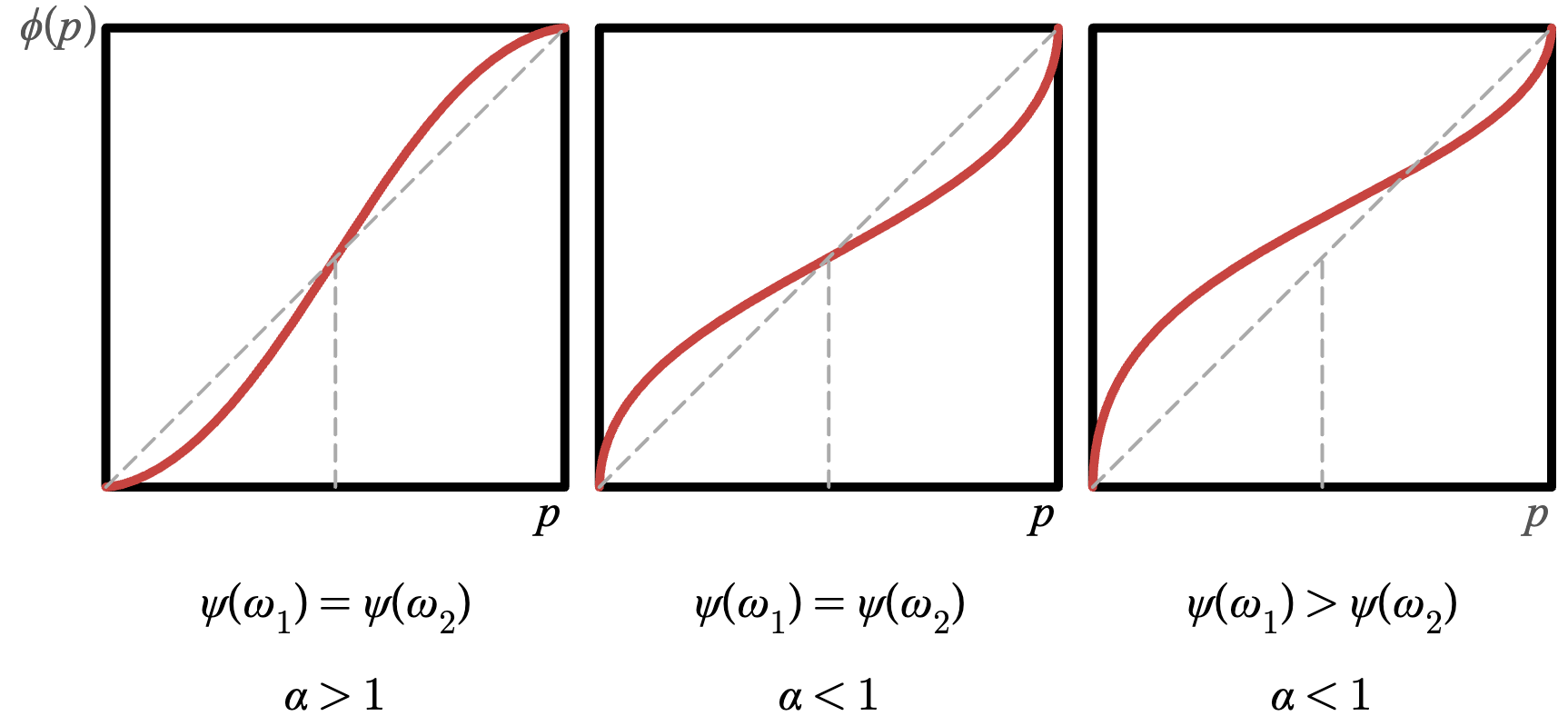}} 
 \caption{\footnotesize{Power distorted function for $\omega_1$ given two states. While $p$ represented the probability of state $1$, $\phi(p)$ represented the distorted belief. The shape of $\phi$ depends on both $\alpha$ and state weights. When $\alpha >1$, it is always $S$-shaped. On the other hand, $\alpha <1$ implies inverse $S$-shaped. The curve intersects the $45^o$ line at $p=0.5$ if the weights of states are equal.  }}
 \label{fig:weight}
\end{figure}

In order to better understand the relationship of our power-weighted distortion functions we now provide a brief exploration of how the parameters of the model affect the shape of the distorted probabilities.  As should come as no surprise, $\alpha$ controls the curvature of the distortion function.  Consider $p^{\alpha}$ for $\alpha>1$: this distorts any $p$  between 0 and 1 to a lower value (and the reverse for $\alpha<0$).  However, because we normalize (in the denominator of the distortion) by the sum of the 
$p'^{\alpha}$ over all state $p'$, the final distortion is not purely convex or concave.  In particular, the distortion function is either S-shaped, or inverse-S-shaped, depending on whether $\alpha>1$ or $<1$ respectively --- see Figure \ref{fig:weight} for an illustration.  Recall that an inverse-$S$ shaped weighting function is often used to capture empirical regularities in the literature on prospect theory-based probability weighting, beginning with \cite{kahneman1979prospect}.\footnote{For additional work on the shape of the weighting function in prospect theory (and cumulative prospect theory), and especially the role of power weighting functions, see \cite{tversky1995weighing,wu1996curvature,gonzalez1999shape,abdellaoui2000parameter,bleichrodt2000parameter,kilka2001determines,diecidue2009parametric}.}  Thus, by controlling $\alpha$ we can effectively control the shape of the weighting function, and for $\alpha<1$ it matches the widely used inverse-S-shaped weighting function.  The additional parameters of the model $\psi(\omega)$, also impact the shape of the distortion function. As Figure \ref{fig:weight} shows with two states, when $\psi$ is constant, the inflection point of the distortion function is equal to $\frac{1}{2}$ (see the middle panel).  But when the weight on state 1 increases, the inflection point of the distortion function for that state shifts upwards --- in other words, probabilities are distorted upwards for a larger set of $p$s.  This should not be surprising, since a larger $\psi$ implies increased overweighting in the distorted probabilities.

When $\alpha=1$, the structure is a bit simpler.  If $\psi(\omega_1) >\psi(\omega_2)$ then the the distortion function for $\omega_1$ is concave, and  strictly above the $45^o$ line (recall that $\phi(0)=0$ and $\phi(1)=1$); and so state 1 is always overweighted.  In contrast, if $\psi(\omega_1)<\psi(\omega_2)$ then the distortion function for $\omega_1$ is convex and lies below the $45^o$ line.  

\section{Transforming Blackwell Matrices}\label{sec:blackwell}
Although many approaches to belief distortions assume that the distortions occur with respect to the probabilities assigned to states, without any reference to potential signals, other models attempt to directly account for the possibility of distortions of observable (and objective) signals. This include some models of conservatism \cite{edwards1968conservatism, augenblick2021overinference}, where individuals underweight the informativeness of signals and models of motivated beliefs such as 
\citet{caplin2019wishful}, \citet{benabou2002self} and \cite{mobius2022managing} where individuals are overly optimistic because they distort the informativeness of signals.\footnote{Models explaining violations of EU are ``static'' and so typically do not consider distortion of signals.}  


We now extend our formal setup to consider signal distortions.  We let $\Theta$ be a finite set, representing the possible outcomes of a signal.  We define a \emph{signal} or \emph{Blackwell experiment} as a map $\sigma:\Omega\rightarrow\Delta(\Theta)$.  The set of Blackwell experiments is denoted by $\Sigma$.  Blackwell experiments are the canonical model of ``noise'' terms for general probability measures, and isolating them from the joint distribution with states allows for many classical comparative static results (e.g., \citet{blackwell, milgrom1981}).  


Given our interest in distorting signals we consider a mapping from a set of probability distributions over signals, with one distribution for each state, to itself.  Technically, our object of interest is a map $f: \Sigma \rightarrow \Sigma$, but we wish to restrict attention to those maps for which for each $\omega\in\Omega$, the distortion $f(\sigma)(\omega)$ depends only on $\sigma(\omega)$.  To this end, take as our primitive a collection of $g_{\omega}:\Delta(\Theta)\rightarrow\Delta(\Theta)$.  Call such an object a \emph{Blackwell distortion function}. 
With a slight abuse of notation denote the distribution over signals, conditional on state $\omega$ (that is, $\sigma(\omega)$) as $\sigma_{\omega}$ (see Figure \ref{fig:blackwellexp}).

\begin{figure}[htbp]
\begin{equation*}
\begin{tabular}{ccccc}
\toprule
& \multicolumn{3}{c}{Signals} & \\ \cmidrule(lr){2-4}
\multicolumn{1}{c}{\multirow{-2}{*}[0.2ex]{$\Omega$}}
 & \ $\theta_1$ \  & $\theta_2$ \  & $\theta_3$  &\multicolumn{1}{c}{\multirow{-2}{*}[0.5ex]{ }} \\ \midrule
\multicolumn{1}{c|}{$\omega_1$} & $\sigma_{\omega_1}(\theta_1)$ & $\sigma_{\omega_1}(\theta_2)$ & $\sigma_{\omega_1}(\theta_3)$ & \multicolumn{1}{|c}{1} \\ \hline
\multicolumn{1}{c|}{$\omega_2$} & $\sigma_{\omega_2}(\theta_1)$ & $\sigma_{\omega_2}(\theta_2)$ & $\sigma_{\omega_2}(\theta_3)$ & \multicolumn{1}{|c}{1} \\ \hline
\multicolumn{1}{c|}{$\omega_3$} & $\sigma_{\omega_3}(\theta_1)$ & $\sigma_{\omega_3}(\theta_2)$ & $\sigma_{\omega_3}(\theta_3)$ & \multicolumn{1}{|c}{1}  \\  
\bottomrule
\end{tabular} \xrightarrow{{\{g_{\omega}\}_{\omega\in\Omega} }} \begin{tabular}{ccccc}
\toprule
& \multicolumn{3}{c}{Signals} & \\ \cmidrule(lr){2-4}
\multicolumn{1}{c}{\multirow{-2}{*}[0.5ex]{$\Omega$}}
 & \ $\theta_1$ \  & $\theta_2$ \  & $\theta_3$  &\multicolumn{1}{c}{\multirow{-2}{*}[0.2ex]{  }} \\ \midrule
\multicolumn{1}{c|}{$\omega_1$} & $g_{\omega_1}(\sigma_{\omega_1})(\theta_1)$ & $g_{\omega_1}(\sigma_{\omega_1})(\theta_2)$ & $g_{\omega_1}(\sigma_{\omega_1})(\theta_3)$ & \multicolumn{1}{|c}{1} \\ \hline
\multicolumn{1}{c|}{$\omega_2$} & $g_{\omega_2}(\sigma_{\omega_1})(\theta_1)$ & $g_{\omega_2}(\sigma_{\omega_1})(\theta_2)$ & $g_{\omega_2}(\sigma_{\omega_1})(\theta_3)$ & \multicolumn{1}{|c}{1} \\ \hline
\multicolumn{1}{c|}{$\omega_3$} & $g_{\omega_3}(\sigma_{\omega_1})(\theta_1)$ & $g_{\omega_3}(\sigma_{\omega_1})(\theta_2)$ & $g_{\omega_3}(\sigma_{\omega_1})(\theta_3)$  & \multicolumn{1}{|c}{1}  \\  
\bottomrule
\end{tabular}
\end{equation*}
    \caption{A Blackwell distortion function $\{g_{\omega}\}_{\omega\in\Omega}$}
    \label{fig:blackwellexp}
\end{figure}

Although relatively general, and consistent with many models of distorted signals (e.g., \cite{caplin2019wishful}) the assumption that we distort signals, conditional on each state, separately, is crucial. Some models, e.g.,  \cite{benabou2002self} assume that distortions occur by ``mixing'' across different rows of the Blackwell experiment; this type of distortion is ruled out by our framework.  We allow for more general distortion technologies in the Section \ref{sec:tsd}.   

We now impose a form of coherency that says the distortions should be independent conditioning on the subset of signals, as well as the subset of states.

\begin{definition}
    $\{g_{\omega}\}_{\omega\in\Omega}$ is \emph{Blackwell signal coherent} if for every $\omega\in\Omega$ and every $S\subseteq \Theta$, $g_{\omega}(\sigma_{\omega}(\cdot |S)) =g_{\omega}(\sigma_{\omega})(\cdot|S)$.
\end{definition}


Note that if we relabel Figure \ref{fig:distortionfunction} by $\theta_i$ instead of $\omega_i$, Figure \ref{fig:distortionfunction} becomes an illustration of the distortion function of  $g_\omega$ for state $\omega$. The interpretation here is similar to that in Section \ref{sec:states}: coherency implies that an individual's distorted likelihoods about signals is independent of whether they learned some signals aren't possible before or after distortion.  Hence, the following corollary is an immediate reinterpretation of Theorem~\ref{thm:motivated}.   


\begin{corollary}\label{prop:blackwellsignal}Suppose that $|\Theta|\geq 3$.  
Let $\{g_{\omega}\}_{\omega\in\Omega}$ be a positive and continuous Blackwell distortion function.  The following are equivalent: 
\begin{enumerate}
    \item $\{g_{\omega}\}_{\omega\in\Omega}$ is Blackwell signal coherent.
\item Each $g_{\omega}$ has a power-weighted distorted belief representation.

\end{enumerate}
\end{corollary}

Our interpretation of the distortion function $g_{\omega}$ follows the same lines as our previous interpretation of $\phi$.  In particular, consider the probability of signals conditional on some state $\omega$.  Each signal realization can have a specific weight attached to it. This signal weight can vary by state.  Fix $\alpha_{\omega}=1$. For a given state $\omega$ it could be that the relative chance of signal $\theta$ compared to $\theta'$ is always increased by some scalar $\frac{\psi_{\omega}(\theta)}{\psi_{\omega}(\theta')}$ (compared to the objective probabilities); but for some other state $\omega'$ the relative chance of the two signals is lower after the distortion.  Similarly, for each state, there can be a power $\alpha_{\omega}$.  To understand its impact, assume $\psi_{\omega}=1$.  For some states it could be that the distorted beliefs are weighted towards more (objectively) likely signals (if $\alpha_{\omega}>1$), while the opposite is true for some other state $\omega'$ (where $\alpha_{\omega'}<1$).  More generally, for small $\alpha_{\omega}$ distorted beliefs are determined only by the signal specific weights $\psi_{\omega}$, while for large $\alpha_{\omega}$, the distorted belief is determined by the objective probabilities, and in particular, the signals with the highest objective probabilities are overweighted after the distortion.  

However, because distorted beliefs about signals, conditional on a state, must sum to one, there are some cross signal restrictions in terms of the likelihood ratios of signal realizations. If the distorted beliefs (relative to objective beliefs) indicate that $\omega$, compared to $\omega'$ is relatively more likely after $\theta$, then there must be some other signal where the opposite is true.  In other words, it cannot be the case that the distorted beliefs cause $\omega$ to be more likely than $\omega'$ after all signals (compared to objective beliefs).    

When $\psi_{\omega} =1$ and $\alpha_{\omega}<1$ such signal distortions are consistent with models of extremeness aversion (\cite{benjamin2019errors}) or underweighting of relatively informative signals and overweighting of relatively noisy signals (\cite{augenblick2021overinference}). If, in contrast $\alpha_{\omega}=1$, then such a model can be consistent with the signal distortions discussed in \cite{caplin2019wishful}.

\section{On Grether's formulation}\label{sec:grether}

We previously discussed how individuals may distort beliefs about states of the world, or about signals.  Of course, in many environments, individuals often have beliefs over both states and signals, and may distort both.  In both this and the following section, we will seek to understand what restrictions coherency imposes in these kinds of situations.  

In this section, we take an approach to modeling such joint distortions which dates to a seminal article on distorted beliefs: \cite{grether1980bayes}.  In line with this approach we will approach signals in this section as they are often discussed in information economics, and how they are treated in Section \ref{sec:blackwell}: as Blackwell experiments.  \cite{grether1980bayes} introduces a general model that allows individuals to distort prior probabilities about states as well as the likelihoods of noisy signals independently, which we use as our jumping off point.  

The use of Blackwell experiments allows us to address a richer set of potential updates than our setup in Section \ref{sec:states}.  We considered what happened when we conditioned on an event, which ruled out a subset of states.  Using the technology of Blackwell experiments we can now explore what happens with information which is less extreme, in that it may change beliefs about the relative likelihood of states without specifically ruling any out.  Our primitive objects are thus prior beliefs over states and a Blackwell experiment, along with a distinct distortion for each.  Formally, assume that $|\Omega| \geq 3$ and $|\Theta|\geq 2$, where  $\Omega$ is interpreted as a set of \emph{states of the world}, whereas $\Theta$ is interpreted as a set of \emph{observable signals}.\footnote{When signals only rule out some states, but provide not additional information, then they essentially act to identify an event, which was the focus of our analysis in Section \ref{sec:states}.}


The Grether model has been generalized in many ways (see \cite{benjamin2019errors} for a recent survey). Like Grether's original proposal, most of these papers assume a particular functional form to both distortions of prior and signal probabilities.  We write a general model of Grether's suggestion as follows.  There are continuous, positive maps:  $f:\Delta(\Omega)\rightarrow\Delta(\Omega)$ and for each $\omega\in\Omega$, $g_{\omega}:\Delta(\Theta)\rightarrow\mathbb{R}_+^{\Theta}$.  The function $f$ transforms priors, and each $g_{\omega}$ transforms the signal distribution conditional on $\omega\in\Omega$ to some distorted noise term.  Notably, the signal transformation is allowed to be state-dependent itself. Let $g$ be the collection of $\{g_\omega \}_{\omega \in \Omega}$.\footnote{Just as discussed in the previous section, this assumption is substantive, and rules out some forms of distortion --- e.g., the signals distortions considered in \cite{benabou2002self} where distortions occur by mixing across rows.  Section \ref{sec:tsd} allows for the entire matrix to be distorted at once (as opposed to each row separately).} 

We first formally define Bayesian  updating in this environment. We define it for any $\sigma$ where $\sum_{\theta} \sigma(\theta|\omega) >0$.  In this case the Bayesian posterior of $p$ given $\theta$ and $\sigma$ is written as: $$\mathcal{B}_{\sigma}(p,\theta)(\omega)=\frac{p(\omega)\sigma(\theta|\omega)}{\sum_{\omega'}p(\omega')\sigma(\theta|\omega')}$$ We now define the Grether update for given two functions $f$ and $g$. The Grether update is a Bayesian update for distorted belief over states $f(p)$ and distorted Blackwell experiment $g \circ \sigma$. Formally, the ($f,g$)-\emph{Grether update} for prior $p$, signal $\sigma:\Omega\rightarrow \Delta(\Theta)$ and $\theta\in\Theta$ is defined as $$\mathcal{B}_{g \circ \sigma }(f(p),\theta)(\omega)=\frac{f(p)(\omega)g_{\omega}(\sigma(\cdot|\omega))(\theta)}{\sum_{\omega'}f(p)(\omega')g_{\omega'}(\sigma(\cdot|\omega')(\theta))}.$$

In Grether's original formulation both $f$ and $g_{\omega}$ were power functions (and $g$ did not depend on the state): $f(p)(\omega)\propto p(\omega)^{\alpha}$ and $g_{\omega}(\sigma)(\theta) = \sigma(\theta)^{\beta}$.

A key assumption in our approach is that the distorted signals need not sum to one.  We do so for two reasons: first, many functional forms used in the literature do not require this; second, this flexibility allows for a much greater set of possible distortions (as we will show later). In the previous sections (and in the following section), we always assumed that distorted beliefs were themselves beliefs --- i.e. the sum of values over the relevant set was always equal to 1.  However, in all of the previous sections (and in the following one), this is without loss of generality.  Because we either distorted a distribution over states, or a distribution over signals conditional on a state, if the distorted beliefs did not sum to one, we could simply divide all value by the sum, and recover probabilities.  In contrast, the assumption has bite here.  This is because we are considering the interaction between multiple ``rows'' of the Blackwell experiment.  In particular, if we force the distorted beliefs in each row to sum to unity, then e.g., distorting the probability of $\theta$ conditional on $\omega$ upwards implies that there is some $\theta'$ whose probability, conditional on $\omega$, must be distorted downwards. This means that distorting the posterior belief about $\omega$ conditional on $\theta$ necessarily implies a distortion about the posterior belief of $\omega$ conditional on $\theta'$.  This is a substantive restriction, and one which does not hold if the distorted beliefs need not sum to 1. We allow for this relaxation only for the distortion of the signals (allowing for more flexibility on the part of state distortions, will, in line with the intuition just provided, not alter our results).

\begin{figure}[htbp]
\centering
\begin{subfigure}{.5\textwidth}
  \centering
  \includegraphics[width=.8\linewidth]{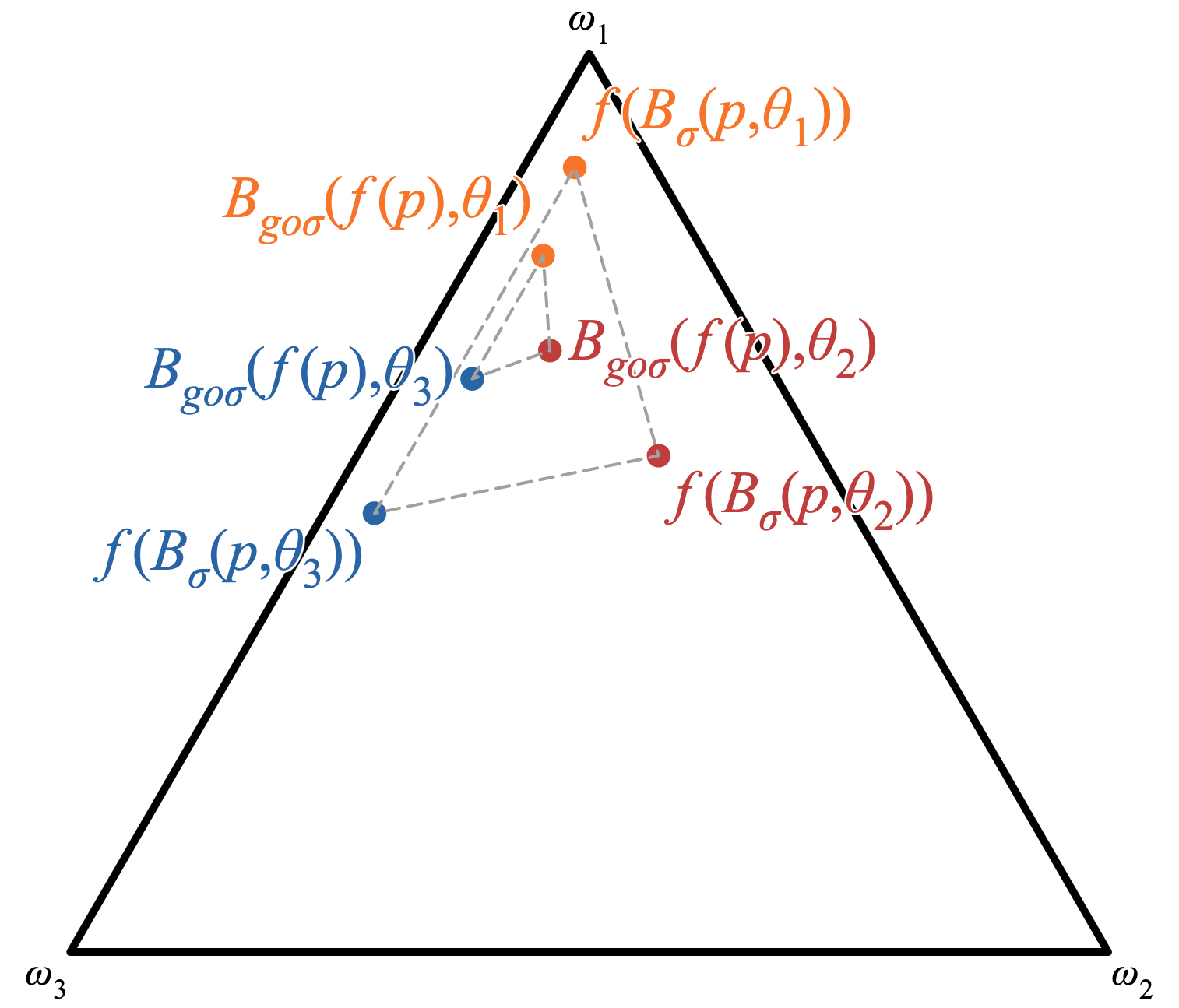}
  \caption{Not Gretherian-coherent}
  \label{fig:grether_1}
\end{subfigure}%
\begin{subfigure}{.5\textwidth}
  \centering
  \includegraphics[width=.8\linewidth]{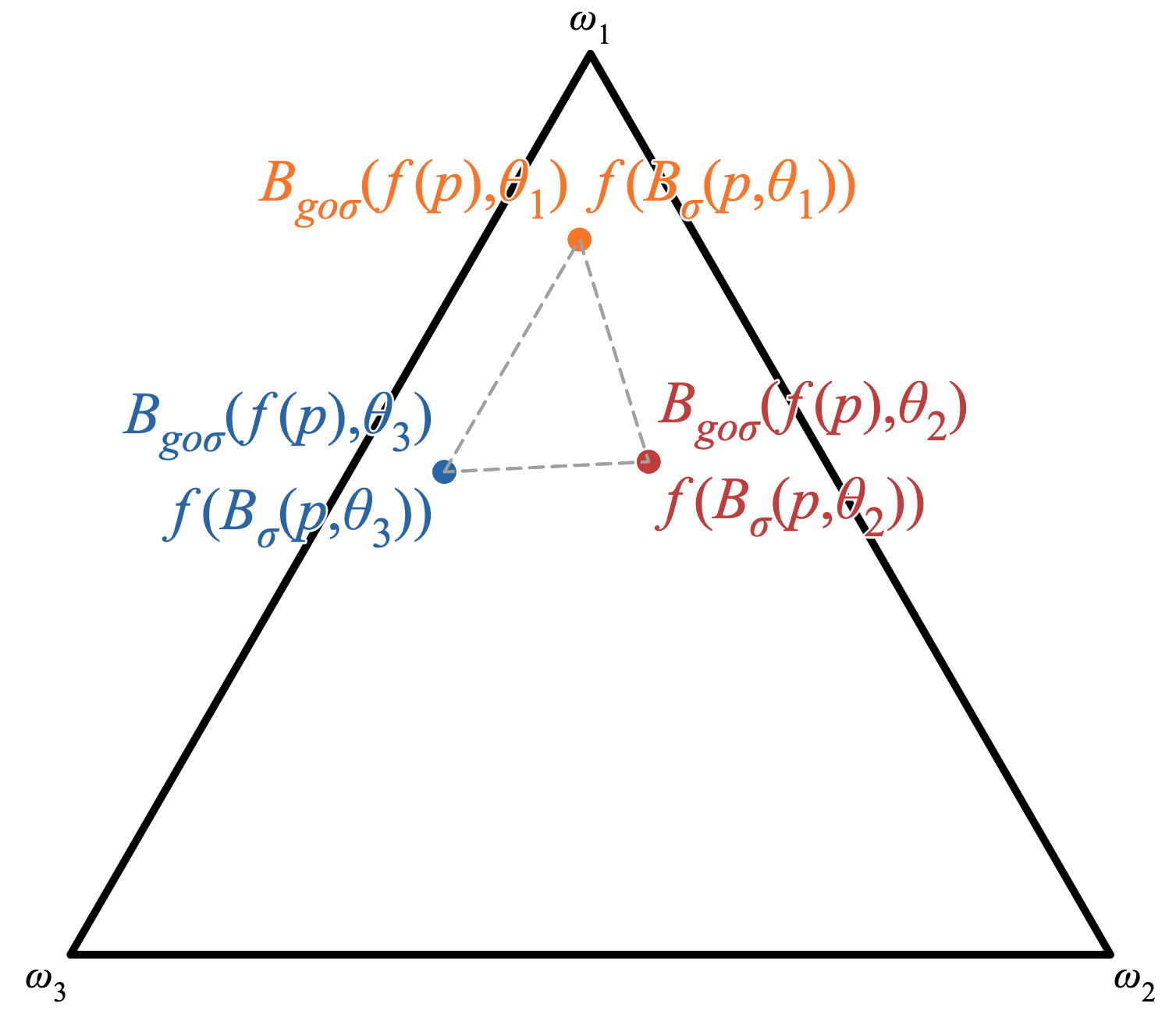}
  \caption{Gretherian-coherent}
  \label{fig:grether_2}
\end{subfigure}
\caption{\footnotesize{This figure illustrates two  Gretherian distortion functions where  $f(p)\propto p^{\alpha}$ and $g_{\omega}(\sigma) = \sigma^{\beta}$. In the left figure, $\alpha$ is different from $\beta$, which implies that it does not satisfy Gretherian coherency. On the other hand, $\alpha$ is equal to $\beta$ in the figure on the right hand. This distortion satisfies Gretherian coherency as illustrated.}}
\label{fig:grether}
\end{figure}

We next turn to understanding what the correct restrictions on the distortion function are required in order to capture coherency.  Recall that coherency requires that the individual's final beliefs be invariant to the ordering of updating and distortions.  In our setting this has important implications.  When the individual distorts and then updates, they apply a distortion to prior beliefs, and a distortion to signals, and then given those distortions, update.  In the latter case, the individual updates using true probabilities, and then distorts.  Notice that in the former, both $f$ and $g_{\omega}$ play a role, since distortions are taken before updating occurs.  In contrast, in the latter, only $f$ is involved.  Formally, we say 
\begin{definition}
     The distorted belief is \emph{Gretherian-coherent} if\begin{equation*}\label{eq:grether}f\left(\mathcal{B}_{\sigma}(p,\theta)\right)(\omega)= \mathcal{B}_{g \circ \sigma }(f(p),\theta)(\omega).\end{equation*} 
\end{definition}
 Of course, this expression is intended to hold when the relevant denominators are positive.  The similarity with Blackwell signal coherence is immediate:  Grether-updating allows apparently additional flexibility in distorting signals when distorting priors before the signal is realized. We provide two examples of Gretherian distortion functions in Figure \ref{fig:grether}. In each figure, there are three possible signals: $\{\theta_1,\theta_2,\theta_3\}$. For each signal, we depict $f\left(\mathcal{B}_{\sigma}(p,\theta_i)\right)$ and $\mathcal{B}_{g \circ \sigma }(f(p),\theta_i)$.  Both  Gretherian distortion functions are of the form of  $f(p)(\omega)\propto  p(\omega)^{\alpha}$ and $g_{\omega}(\sigma) = \sigma^{\beta}$. In the left figure, $\alpha$ is different from $\beta$, which implies that it does not satisfy Gretherian coherency ($f\left(\mathcal{B}_{\sigma}(p,\theta_i)\right) \neq \mathcal{B}_{g \circ \sigma }(f(p),\theta_i)$). In the right figure, $\alpha$ equals $\beta$, hence it satisfies Gretherian coherency as illustrated.

Gretherian-coherency is a strong condition.  Not only must state distortions commute with updating, but they must also be invariant to whatever kind of signal distortions occur.  Thus, there is an apparent tension between the distorted signals and undistorted signals, which will be formalized in Theorem~\ref{thm:grether}.

The next result says that if we impose Grether coherence, distortions again must take a very specific form. Both probabilities and noise terms can be distorted in a power form, but the power must be the same for all relevant variables.

\begin{theorem}\label{thm:grether}Suppose that $|\Omega|\geq 3$ and $|\Theta|\geq 2$.  Suppose $f$ and $g_{\omega}$ for all $\omega\in\Omega$ are given.  Then $f$ and $g_{\omega}$ are positive and continuous and the pair $(f,g)$ is Gretherian-coherent
iff there exists $\psi(\omega)>0$ for all $\omega$, $\gamma(\theta)>0$ for all $\theta$, and $\alpha > 0$ for which $f(p)(\omega)=\frac{\psi(\omega)p(\omega)^{\alpha}}{\sum_{\omega'}\psi(\omega')p(\omega')^{\alpha}}$ and $g_{\omega}(\sigma(\cdot|\omega))(\theta)=\gamma(\theta)(\sigma(\theta|\omega))^{\alpha}$ for all $\omega$.\end{theorem} 

Grether coherency is stronger in two ways than our previous definition of coherency.  As noted previously, our original notion of coherency only applied when thinking about the realization of a particular event.  In other words, information only had the role of ruling out states.  In contrast, Grether coherency applies after any kind of information.  Second, our previous notion of coherency did not allow distortions regarding what could be learned from event: if we think of the possible events that could be learned as coming from a Blackwell experiment, we previously assumed that the entries in that experiment could not be distorted.  In contrast, in this section we do allow for distortions regarding the informational content of a signal.  
Although the function form restrictions for state distortions are not more restrictive than previously, we have relatively strong restrictions on the form of the signal distortions.  First, here, unlike in Section \ref{sec:blackwell}, the signal specific weights $\gamma(\theta)$ must be state-\emph{independent}.  Previously they could vary by state.  This means that if, in state $\omega$, signal $\theta$ is overweighted relative to the objective probability, it must also be true in state $\omega'$.  Second, the power $\alpha$ is the same for both signal and state distortions.  For example, if the distorted state probabilities put more the weight on the objectively most likely state (due to a large $\alpha$), then the distorted signal probabilities must put more  weight on the objectively most likely signal as well.  Conversely, if the distorted probabilities over states are independent of the objective probabilities, then the same must be true for the distorted probabilities over signals.  Third, because the distorted signal probabilities do not need to sum to 1, it is no longer the case (unlike in Section \ref{sec:blackwell}) that making the likelihood of state $\omega$ relative to $\omega'$, conditional on signal $\theta$, higher due to a distortion means that there must be some other signal where the opposite it true.  

Gretherian coherency is consistent with several classes of models of distorted beliefs.  For example, if $\psi=1$ and $\alpha<1$ such a a model allows for both base rate neglect and conservatism, but with the restriction that the DM must exhibit the ``same'' degree of base-rate neglect as conservatism.  Similarly if $\psi=1$ but $\alpha>1$ then the DM must exhibit excess sensitivity to base-rates and overreaction to signals.  Alternatively, setting $\alpha=1$, we can instead generate distortions associated with a model of motivated beliefs where both priors and signal probabilities are distorted.  

Observe that by Theorem~\ref{thm:grether}, $f$ is coherent as discussed in Section~\ref{sec:states}.  Further, up to normalization, $g_{\omega}$ is Blackwell signal coherent in the sense of Section~\ref{sec:blackwell}.  Hence, imposing either of these as additional conditions adds nothing.  One caveat is that in order for $g_{\omega}$ to be formally Blackwell signal coherent, it needs to map probabilities to probabilities.  This is addressed in the following corollary.

The following corollary demonstrates the reasoning behind permitting signal distortions to not sum to one:  such a restriction automatically implies that $\alpha = 1$.  A corollary of this result is a finite-states version of a result due to \citet{genest}, which implicitly presumes each $g_{\omega}$ is the identity map (see Theorem 2.2 of that paper).

\begin{corollary}\label{cor:grether2}Suppose, in addition to the hypotheses of Theorem~\ref{thm:grether}, that for each $\omega$ and each $\sigma$, $\sum_{\theta}g_{\omega}(\sigma(\cdot|\omega))(\theta)=1$.  Then there exists $\psi(\omega)> 0$ for each $\omega$, for which $f(p)(\omega)=\frac{\psi(\omega)p(\omega)}{\sum_{\omega'}\psi(\omega')p(\omega')}$ and each $g_{\omega}$ is the identity map.\end{corollary}

Here, it must be the case that the distortions to the probabilities of states take the weighted form.  Relative to the results of Theorem~\ref{thm:grether}, the additional restriction that distorted signal probabilities must sum to one imposes substantive restrictions.  In particular, no distortion of signal probabilities are allowed, and distortion of state probabilities can only be weighted.  For example, models of base-rate neglect, such as those originally proposed by \cite{grether1980bayes} and \cite{benjamin2019base}, are not allowed, nor are models of under-inference. Similarly, motivated beliefs about states are allowed (but not motivated beliefs regarding signals).

\section{Transforming signal distributions}\label{sec:tsd}

In this section we approach combining distortions of signals and of states in a distinct way from the previous section.  Recall that in Section \ref{sec:grether} the object of study was the combination of a set of prior beliefs and a Blackwell experiment.  In contrast, here we follow the representation of \citet{green2022}, and assume that probabilistic uncertainty is over $\Omega\times\Theta$.  This framework allows us to investigate what happens upon observing a signal.  As before, our interest is in an environment where signals are observed, either before or after distorting beliefs, and imposing consistency requirements across these different timings.

Each $p\in\Delta(\Omega\times\Theta)$ specifies a joint distribution over states and signals:  the probability over all pairs $(\omega,\theta)\in\Omega\times\Theta$ must sum to one.  We think of this joint distribution as being \emph{objective} and hence subject to distortion in the same way a distribution over states is.

We can represent these joint distributions in matrix form, which we call GS matrices (for Green and Stokey).  Notice that the sum of the elements of a GS matrix must sum to one.  This joint distribution, and the associated GS matrix, should be distinguished from a Markov matrix or as a Blackwell experiment (\citet{blackwell}). If one were given a Blackwell experiment and a prior over states, the GS matrix is the induced distribution over state, signal combinations that would result.   Similarly, given a GS matrix, one can obtain the prior over states by summing the elements of any given row.  After having done so, one can then divide the values in each row by the sum of the elements in the row, so that the sum in each row is equal to 1 (see the left side of Figure \ref{fig:jointdistribution}).  This then generates a Blackwell experiment.  


Thus, in this section, we explore what happens when individuals distort the GS matrix, so that distorted beliefs over states end up being the row marginals, and the distorted Blackwell matrix can be recovered via the normalization discussed above. In the last section, we formally investigate what happens when the DM can distort the prior and the Blackwell experiment directly. In this section, we consider a general framework whereby decision-makers are permitted to ``distort'' the probabilities associated with the realizations of signals. This allows us to distort the joint probabilities at the same time.


We will study what happens upon learning a signal (or subset of signals).  Recall  $S \subseteq \Theta$, is a subset of signals (i.e. rows), and we will call it a signal-event (a singleton signal event is just a signal).\footnote{As a reminder, signals which only rule out some states, but provide no further information, act to identify events, the unit of analysis in Section \ref{sec:states}.}

To this end, consider our distortion function $\phi:\Delta(\Omega\times\Theta)\rightarrow\Delta(\Omega\times\Theta)$.  This map specifies the distorted \emph{joint} distribution over states and signals.  We can now define the appropriate notion of coherency.\footnote{We continue to use the same definitions of positivity and continuity as before.} To simplify on notation, we often write $\omega$ to refer to the event $\{\omega\}\times\Theta$, so \emph{e.g.} $p(\omega)$ means $p(\{\omega\}\times\Theta)$.  In a similar fashion, $p(\theta)$ refers to $p(\Omega\times\{\theta\})$.  Likewise, $p(\cdot|\theta)=p(\cdot|\Omega\times \{\theta\})$.  We often drop $\{\}$ brackets to simplify the exposition.

\begin{definition}
 $\phi$ is \emph{weakly distorted signal coherent} if for all $\theta\in\Theta$ for which $p(\theta)>0$, we have $\phi(p(\cdot|\theta))=\phi(p)(\cdot|\theta)$. $\phi$ is \emph{strongly distorted signal coherent} if for all signal events $S \subseteq \Theta$, for which $p(\Omega\times S)>0$, we have  $\phi(p(\cdot|S))=\phi(p)(\cdot|S)$. 
\end{definition}


While strong signal coherence requires that the distortion commute with respect to updating of any subset of signals (i.e. any signal event), weak signal coherence only requires commutativity with respect to updating upon a single signal.  We first interpret weak distorted signal coherency.  Notice that this is a natural analogue to coherency which was discussed Section \ref{sec:states}.  Recall that conditioning on a particular $\theta$ (i.e. being in a particular column in the GS matrix) is the same as the DM observing a signal and conditioning their beliefs on the realization of that signal. Thus, coherency asks, given any realized signal, that the agent must have the same final beliefs if they first distort, then update given the signal, or if instead they update given the signal and then distort. Alternatively, a second interpretation is simply that the decision-maker learned that the only relevant column in the GS matrix (i.e. the only column that is actually possible) is $\theta$. Strong distorted signal coherency is similar, but now the conditioning is instead done on knowing any subset of signals is possible (rather than just one, as in weak distorted signal coherency). With $n$ signals, weak signal coherence only imposes restrictions on at most $n$ different updated probabilities. On the other hand, strong signal coherence imposes conditions on up to $2^n-2$ updated probabilities. However, we will ultimately show that the distinction between the two is not so large. 

These definitions are based on the idea that signals are observed before the state of the world is realized.  Of course, despite the similarities, our definition differs from that of coherency utilized in Section \ref{sec:states} in an important way.  There the conditioning events were subsets of the state space.  In contrast, here, the event is a subset of signals (or, in the case of weak coherence, a single signal). Thus, states themselves are not directly observed, though if $p\in\Delta(\Omega\times\Theta)$ is such that $p(\omega|\theta)=1$ for some pair $(\omega,\theta)\in\Omega\times\Theta$, then once one learns $\theta$, one infers immediately that $\omega$ obtained. %

\begin{figure}[htbp]
\begin{equation*}
\begin{tabular}{ccccc}
\toprule
& \multicolumn{3}{c}{Signals} & \\ \cmidrule(lr){2-4}
\multicolumn{1}{c}{\multirow{-2}{*}[0.5ex]{$\Omega$}}
 & \ $\theta_1$ \  & $\theta_2$ \  & $\theta_3$ &\multicolumn{1}{c}{\multirow{-2}{*}[0.5ex]{ Prior}} \\ \midrule
\multicolumn{1}{c|}{$\omega_1$} & $p_{11}$ & $p_{12}$ & $p_{13}$ &  \multicolumn{1}{|c}{$p(w_1)$}\\ \hline
\multicolumn{1}{c|}{$\omega_2$} & $p_{21}$ & $p_{22}$ & $p_{23}$ & \multicolumn{1}{|c}{$p(w_2)$} \\ \hline
\multicolumn{1}{c|}{$\omega_3$} & $p_{31}$ & $p_{32}$ & $p_{33}$ & \multicolumn{1}{|c}{$p(w_3)$}\\  
\cmidrule(lr){2-4}\cmidrule(lr){4-5}
 & $p(\theta_1)$ & $p(\theta_2)$ & $p(\theta_3)$ & \multicolumn{1}{|c}{1} \\
 \bottomrule
\end{tabular} \xrightarrow{{\ \ \phi \ \ \ }} \begin{tabular}{ccccc}
\toprule
& \multicolumn{3}{c}{Signals} & \\ \cmidrule(lr){2-4}
\multicolumn{1}{c}{\multirow{-2}{*}[0.5ex]{$\Omega$}}
 & \ $\theta_1$ \  & $\theta_2$  & $\theta_3$ &\multicolumn{1}{c}{\multirow{-2}{*}[0.5ex]{ Prior}} \\ \midrule
\multicolumn{1}{c|}{$\omega_1$} & $q_{11}$ & $q_{12}$& $q_{13}$  & \multicolumn{1}{|c}{$q(w_1)$}\\ \hline
\multicolumn{1}{c|}{$\omega_2$} & $q_{21}$ & $q_{22}$ & $q_{23}$ & \multicolumn{1}{|c}{$q(w_2)$} \\ \hline
\multicolumn{1}{c|}{$\omega_3$} & $q_{31}$ & $q_{32}$ & $q_{33}$ & \multicolumn{1}{|c}{$q(w_3)$}\\ 
\cmidrule(lr){2-5}
 & $q(\theta_1)$ & $q(\theta_2)$ & $q(\theta_3)$ & \multicolumn{1}{|c}{1}  \\
  \bottomrule
\end{tabular}
\end{equation*}

    \caption{A distortion function over a joint distribution over states and signal}
    \label{fig:jointdistribution}
\end{figure}

Figure \ref{fig:jointdistribution} shows an example of a joint distribution matrix $P$.  Notice that $p(\omega_i) = \sum_j p_{ij}$, and $p(\theta_j) = \sum_i p_{ij}$.  We then consider a distortion, resulting in the joint distribution matrix $Q$.  We define $q_{ij} = \phi(p_{ij})$, while $q(\omega_i) = \sum_j q_{ij}$, and $q(\theta_j) = \sum_i q_{ij}$.  Thus, notice that in contrast to the previous section, we do not directly distort the probabilities of states.  Distortions of states occur through the distortion of the joint distribution matrix.
In other words, distorted entries in the experiment then lead to distorted prior beliefs and probabilities of observing a signal.

To consider the implications of coherency, we think about what happens if we first condition on a subset of the signals.  Let signal event $E_{12}$ be the set of signals $\theta_1$ and $\theta_2$, highlighted in red on the left panel of Figure \ref{fig:blackwellcondition}.  We call the probability of this event  $p_{11}+p_{12}+p_{21}+p_{22}+p_{31}+p_{32}$.  We can then consider what the new Blackwell experiment looks like, which is $P$ conditional on $E_{12}$, denote this $P_{E_{12}}$, as shown in the left panel of Figure \ref{fig:blackwellcondition} where $p'_{ij}=\frac{p_{ij}}{p_{11}+p_{12}+p_{21}+p_{22}+p_{31}+p_{32}}$.  Notice we can simply ignore the third column in the experiment at this point since we are conditioning on the first two columns.  We can then apply the distortion function $\phi$ to $P_{E_{12}}$, which generates a new distorted matrix $Q_{E_{12}}$. Our coherency condition relates $Q$ and $Q_{E_{12}}$:  it says that $$q'_{11}=\frac{q_{11}}{q_{11}+q_{12}+q_{21}+q_{22}+q_{31}+q_{32}}$$


\begin{figure}[htb]
\begin{equation*}
\begin{tabular}{ccccc}
\toprule
& \multicolumn{3}{c}{Signals} & \\ \cmidrule(lr){2-4}
\multicolumn{1}{c}{\multirow{-2}{*}[0.5ex]{$\Omega$}}
 & \ $\theta_1$ \  & $\theta_2$ \  & $\theta_3$ &\multicolumn{1}{c}{\multirow{-2}{*}[0.5ex]{ Prior}} \\ \midrule
\multicolumn{1}{c|}{$\omega_1$} & $\tikzmarkin{c}p'_{11}$ & $p'_{12}$ & - &  \multicolumn{1}{|c}{$p'(\omega_1)$}\\ \hline
\multicolumn{1}{c|}{$\omega_2$} & $p'_{21}$ & $p'_{22}$& - & \multicolumn{1}{|c}{$p'(\omega_2)$}\\ \hline
\multicolumn{1}{c|}{$\omega_3$} & $p'_{31}$ & $p'_{32}\tikzmarkend{c}$  & - & \multicolumn{1}{|c}{$p'(\omega_3)$}\\ 
\cmidrule(lr){2-4}\cmidrule(lr){4-5}
 & $p'(\theta_1)$ & $p'(\theta_2)$ & - & \multicolumn{1}{|c}{1} \\
\bottomrule
\end{tabular} \xrightarrow{{\ \ \phi \ \ \ }} \begin{tabular}{ccccc}
\toprule
& \multicolumn{3}{c}{Signals} & \\ \cmidrule(lr){2-4}
\multicolumn{1}{c}{\multirow{-2}{*}[0.5ex]{$\Omega$}}
 & \ $\theta_1$ \  & $\theta_2$ \  & $\theta_3$ &\multicolumn{1}{c}{\multirow{-2}{*}[0.5ex]{ Prior}} \\ \midrule
\multicolumn{1}{c|}{$\omega_1$} & $q'_{11}$ & $q'_{12}$& -  & \multicolumn{1}{|c}{$q'(s_1)$}\\ \hline
\multicolumn{1}{c|}{$\omega_2$} & $q'_{21}$ & $q'_{22}$ & - & \multicolumn{1}{|c}{$q'(s_2)$} \\ \hline
\multicolumn{1}{c|}{$\omega_3$} & $q'_{31}$ & $q'_{32}$ & - & \multicolumn{1}{|c}{$q'(s_3)$}\\
\cmidrule(lr){2-4}\cmidrule(lr){4-5}
 & $q'(\theta_1)$ & $q'(\theta_2)$ & - & \multicolumn{1}{|c}{1} \\
\bottomrule
\end{tabular}
 \end{equation*}
    \caption{A distortion function, conditioned on a signal event $E_{12}$}
    \label{fig:blackwellcondition}
\end{figure}

Given these definitions, we can now describe distorted signal coherent $\phi$.  The following proposition is stated without proof, though the first part (the characterization of weak signal coherence) appears in the proof of the following Theorem~\ref{thm:verygeneral}.

We use the notation $p(\cdot;\theta)$ to denote the probability distribution in $\Delta(\Omega)$ for which $p(\omega;\theta)=p(\omega|\theta)$.  So, while $p(\cdot|\theta)$ is technically an element of $\Delta(\Omega\times\Theta)$ where most probabilities are zero, $p(\cdot;\theta)$ ignores these zero probability events.


\begin{remark}\label{rem:coherence}Weak signal coherence on itself is not particularly strong.  Suppose that $\Omega$ and $\Theta$ are finite.  An exhaustive method for constructing weakly signal coherent, continuous and positive distortions is as follows.  First, fix some continuous $\varphi:\Delta(\Omega\times\Theta)\rightarrow\Delta(\Theta)$ for which $p(\theta)>0$ implies $\varphi(p)(\theta)$.  This can be understood as the distortion of signal probabilities; notably it is allowed to depend on the entire distribution over $\Omega\times\Theta$.  For each signal $\theta\in\Theta$, let $h_{\theta}:\Delta(\Omega)\rightarrow\Delta(\Omega)$ be a positive and continuous function.  Define $$\phi(p)(\omega,\theta)=\varphi(p)(\theta)h_{\theta}(p(\cdot;\theta))(\omega).$$  Then $\phi$ satisfies weak signal coherence, positivity and continuity.  Implicit in Theorem~\ref{thm:verygeneral} below is the fact that any such function can be represented in this way.  Strong signal coherence adds some bite, but imposes structure only on the function $\varphi$.  It is consistent with the idea that there is, for each $\theta\in\Theta$, a continuous $\gamma_{\theta}:\Delta(\Omega)\rightarrow\mathbb{R}_{++}$ and $\alpha > 0$ for which $\varphi(p)(\theta)=\frac{\gamma_{\theta}(p(\cdot;\theta))p(\theta)^{\alpha}}{\sum_{\theta'\in\Theta}\gamma_{\theta'}(p(\cdot;\theta'))p(\theta')^{\alpha}}$. In fact we believe this family to exhaust all distortions satisfying strong signal coherence, though a proof of this is beyond the scope of this work.\end{remark}

According to Remark~\ref{rem:coherence}, and perhaps surprisingly, the set of such distortion functions is large -- in other words, either form of signal coherency appears to be a relatively weak condition on distortion functions.  However, this flexibility should not be too surprising:  weak signal coherency is a much weaker condition than Gretherian consistency.  Indeed, weak signal coherency only requires conditionining on singleton signals.  Recall that Gretherian consistency imposed that distorting both prior beliefs and the Blackwell matrix, and then doing Bayesian updating must be equivalent to doing first Bayesian updating and then distorting the posterior beliefs.  However, distorted signal coherency allows us to use the information contained in the signals (i.e. the entries in each row) rather than just working with the row marginals.

In some sense, the focus on beliefs over states that Gretherian consistency imposes is natural for economists. Ultimately, we care about state distortions, rather than signal distortions; even though we allow the DM the flexibility to transform signals.  This is because we suppose the DMs payoffs depend on the action taken and the state, the signals are merely ancillary in that they can change beliefs about the state. The information about states is much coarser than that contained in the GS matrix.  In fact, beliefs over states, as discussed, are the row marginals of the GS matrix.   

Thus, in order to achieve distortions that allow us to work only on posterior states of the world, we need to impose an additional condition:  whether a given model of distortions of the GS matrix can induce consistent distortions of beliefs over states? In other words, if we are only concerned about distortions of the probabilities of states, can we define a ``reduced form'' distortion operator that acts only on those, but which is consistent with the distortion of the GS matrix. In doing so we ask that our distortion function allow us to discern the distorted state probabilities independently of what the underlying distortion probabilities of state-signal pairs are.  To this end, we provide an additional definition.

\begin{definition}
    A distortion function $\phi$ satisfies \emph{marginality} if for any $p,p'\in\Delta(\Omega\times\Theta)$, if for all $\omega\in\Omega$, $p(\omega)=\sum_{\theta\in\Theta}p(\omega,\theta)=p'(\omega)$, then for all $\omega\in\Omega$, $\phi(p)(\omega)=\phi(p')(\omega)$.  
\end{definition}

Put differently, for any $p\in\Delta(\Omega\times\Theta)$, we denote by $p|_{\Omega}\in\Delta(\Omega)$ the marginal of $p$ on $\Omega$, so that $p|_{\Omega}(\omega)=p(\omega)=\sum_{\theta\in\Theta}p(\omega,\theta)$ for every $\omega\in \Omega$.  Then marginality is the requirement that $p|_{\Omega}=p'|_{\Omega}$ implies $\phi(p)|_{\Omega}=\phi(p')|_{\Omega}$.  This captures the fact that although state probabilities are derived as the marginals from the GS matrix, we can capture the impact of distortions on the state probabilities by only distorting those marginals (rather than having to consider the distortion of the entire GS matrix).

Marginality allows us to meaningfully define a distortion of state probabilities $\phi^M:\Delta(\Omega)\rightarrow\Delta(\Omega)$ via $\phi^M(q)=\phi(p)|_{\Omega}$ for any $p\in\Delta(\Omega\times\Sigma)$ for which $p|_{\Omega}=q$.  Marginality on its own is relatively unrestrictive, in the same sense that weak signal coherence is unrestrictive.  Though a full characterization is not particularly illuminating, one could be constructed along the lines of Remark~\ref{rem:coherence}.

Our main result constitutes a characterization of distortion functions satisfying both of these properties:  weak signal coherence and marginality.  It establishes that their conjunction is quite strong, even though each of them is individually relatively innocuous.


\begin{theorem}\label{thm:verygeneral}Suppose that $|\Omega|\geq 3$ and $|\Theta|\geq 2$.  Suppose further that $\phi:\Delta(\Omega\times\Theta)\rightarrow\Delta(\Omega\times\Theta)$ satisfies positivity, continuity, weak  signal coherence, and marginality.  Then for each $\omega\in\Omega$, there is $\psi(\omega)>0$ for which for all $q\in\Delta(\Omega)$, $\phi^M(q)(\omega)=\frac{\psi(\omega)q(\omega)}{\sum_{\omega'}\psi(\omega')q(\omega')}$.  Further, for all $\theta\in\Theta$ we have $\phi(p)(\omega,\theta|\{\theta\})=\frac{\psi(\omega)p(\omega,\theta)}{\sum_{\omega'}\psi(\omega')p(\omega',\theta)}$ when $p(\theta)>0$.  Finally, under either
\begin{enumerate}
\item $|\Omega|\geq |\Theta|$, or
\item $|\Theta|\geq 3$ and $\phi$ is strongly signal coherent
\end{enumerate} we additionally have $$\phi(p)(\omega,\theta)=\frac{ \psi(\omega)p(\omega,\theta)}{\sum\limits_{\omega',\theta'}\psi(\omega')p(\omega',\theta')}.$$
\end{theorem}


The theorem has quite strong implications: it shows that the combination of coherency and marginality significantly restricts the set of state-distortions that are possible: they must be power-weighted but where the weight $\alpha$ is always equal to 1. In other words, each signal-state pair can be weighted, where the weight is determined by the state.  Thus, if one signal, conditional on a state, is overweighted, all signals, conditional on the same state,  must be overweighted by the same amount.  Thus, the weight applied to the probability of that specific state (which is simply the row marginal of the GS matrix) must also be overweighted by the same amount.  Moreover, if we map the distorted GS matrix into the equivalent prior plus Blackwell experiment, this implies that there must be no distortions at all to the Blackwell experiment.  

Such a result is stronger than our main result under Gretherian consistency: Theorem~\ref{thm:grether}, which allowed for signals and state probabilities to also be raised to a power under distortions (albeit the same power).  However, the conclusion is essentially the same as the more restrictive result we obtained there, Corollary \ref{cor:grether2}, which is what was obtained when we imposed not just Gretherian consistency but also that the signal distortions must map to probability distributions. Thus, although in principle it seems much more general to allow for the distortion of a full matrix (as we do here), rather than row-by-row (as in the previous section), it ends up not being.  

In fact, this formulation rules out many forms of biases allowed by our Gretherian consistency, including Grether's original model, as well as any model that imposes signal distortions.  However, it still allows for forms of biases where states are over or underweighted, as in many models of motivated beliefs.

\section{Relationships}\label{sec:apps}
We now turn to relating our results to existing approaches in several settings.  In most of these, we try to understand how our belief distortions relate to well-known models of preferences.  In order to do so, we need to make assumptions about how an individual integrates beliefs with values assigned to outcomes to generate preferences.  We begin by assuming that an individual maximizes expected utility, given their distorted beliefs.  We discuss how to apply our model in the world of risky real-valued payoffs, and show that imposing continuity implies the coherent distortions must be signal coherent, and that induced choices can be described by Chew's weighted utility model. 

In the next subsection, we consider non-expected utility criterion, and show that coherent distortions of states are the natural outcome of models of motivated beliefs where the decision-maker faces a cost of distortion is a variant of the Kullback-Liebler divergence of true and distorted beliefs.  

We then relate our notion of coherent beliefs to dynamically consistent behavior, showing that coherency implies dynamic consistency when the decision-maker is an expected utility maximizer conditional on their beliefs.  
Last, we discuss limit beliefs of coherent distortion function and show that they depend crucially on the value of $\alpha$.

\subsection{Belief-Conditional Expected Utility}\label{sec:eu}
One natural way to induce a preference ordering over risky outcomes, given our belief distortions, is to assume that individuals maximize the expected utility, conditional on the distorted beliefs, when choosing prospects.  Such an assumption has been used in many situations (e.g., \cite{brunnermeier2005optimal} in the domain of motivated beliefs, models of rank-dependent utility and prospect theory {\cite{quiggin1982theory,kahneman1979prospect,tversky1992advances}, and most models applying Grether type models of probabilistic distortions, as in \cite{benjamin2019base}, \cite{benjamin2019errors}, \cite{augenblick2021overinference}, \cite{enke2023cognitive}).   Of course, because individuals distort their beliefs, even though they are an expected utility maximizer conditional on beliefs, their choices may not be consistent with the expected utility maximization. 
Many models of distorted beliefs consider not just probability distributions over state spaces, but also over lottery outcomes.  Thus, in order to examine these preferences we must extend our approach to allow for this.   Formally, assume that there exists a set of monetary outcomes $X =[\underline{x}, \bar{x}]\subseteq\mathbb{R}$, where $\underline{x}<\bar{x}$.  We use the notation $\Delta(X)$ to represent the set of \emph{simple} lotteries (\emph{i.e.} lotteries with finite support).  

For each lottery, and $x \in X$ such that $p(x)>0$ we associate $x$ with a state.  A distortion function then maps a probability vector to another probability vector which has the same support.  In other words $\phi$ maps $\Delta(X)$ to $\Delta(X)$.  Positivity means that the distorted lottery $\phi(p)$ has the exact same support as $p$ --- $\phi(p)(x) =0$ if and only if $p(x)=0$. 

Coherence has much the same interpretation as before.  But in this context, we can meaningfully talk about two lotteries as being ``close'' to one another in a stronger sense than we could in previous sections.  In particular, we use the standard notion of weak convergence.  For a sequence of lotteries $\{p_n\}\subseteq \Delta(X)$, say that $p_n \xrightarrow{\textit{ w }} p^*\in\Delta(X)$ if for every bounded continuous $f:X\rightarrow\mathbb{R}$, we have $\int_Xf(x)dp_n(x) \rightarrow \int_Xf(x)dp^*(x)$.  With this notion, degenerate lotteries on $x$ and $x+\epsilon$ become closer the smaller $\epsilon$ becomes. We then say a distorted belief $\phi$ is \emph{weakly continuous} if $p_n \xrightarrow{\textit{ w }} p^*$ implies $\phi(p_n) \xrightarrow{\textit{ w }}\phi(p^*)$. 

It turns out that weak continuity of a coherent distorted belief requires that $\alpha = 1$: Theorem~\ref{thm:weakcont} demonstrates this formally.   

\begin{theorem}\label{thm:weakcont}A distorted belief $\phi$ on $\Delta(X)$ is positive, coherent, and weakly continuous if and only if there exists a continuous $\psi:X\rightarrow\mathbb{R}_{++}$ for which for all $p\in\Delta(X)$, $\phi(p)(x) = \frac{\psi(x)p(x)}{\sum\limits_{x':p(x')>0}\psi(x')p(x')}$.\end{theorem}

We can then consider preferences over the set of lotteries which are induced by an individual being an expected utility maximizer, when they evaluate an objective lottery $p$ subjectively as if it were $\phi(p)$.  For an individual with utility index $u:X\rightarrow\mathbb{R}$, denote the induced preferences as $\succeq_{u, \phi}$, so that $p \succeq_{u,\phi}q$ if and only if $\int_X u(x) d\phi(p)(x)\geq \int_Xud\phi(q)(x)$.

Although $\succeq_{u, \phi}$ represents expected utility preferences conditional on the distortion function $\phi$, it will not typically satisfy the independence axiom.  

Recall the weighted utility representation of \cite{chew1983generalization}.\footnote{See also \citet{fishburn1983transitive,fishburn1988transitivity,chew1994projective} as well as \citet{bolker} for an earlier related result on a different domain.}  The utility of a lottery $p$ is equal to 

$$\sum_{i} \frac{\psi(x_i) p(x_i)}{\sum_j \psi(x_j) p(x_j)} u(x_i) $$ where $\psi$ is a weighting function, and $u$ is a utility index.  As \cite{chew1983generalization} shows, this model can accommodate well-known violations of expected utility such as the Allais paradox.  Moreover, observe that the set of weighted utility functions is equivalent to the set of expected utility functionals where probabilities are given by a weighted probability distortion.


Thus, a well-known generalization of expected utility naturally captures coherent distorted beliefs.  Moreover, the only suitably continuous coherent distorted beliefs that induce expected utility preferences are those that are weighted. 

\subsection{Belief-Conditional Non-Expected Utility}\label{sec:motivated}

In the previous subsection, we considered preferences where an individual, conditional on their distorted beliefs, maximizes the expected utility.  Of course, it may be the case that even after we condition on an individual's distorted beliefs, their preferences still do not satisfy the expected utility.  We turn to considering such a formulation here, focusing on the case when preferences are defined over states.  

Recall that in the previous subsection, we showed that our model could be derived from an optimization problem where the individual maximizes (or minimizes) the expected utility, conditional on the ``distance'' between true and distorted beliefs, as measured by a generalized version of the Kullback-Liebler (KL) divergence, is less than some amount.  

A distinct way of modeling motivated beliefs, which is widely used in the literature (see  \cite{bracha2012affective}, \cite{mayraz2019priors} and \cite{caplin2019wishful} among others) is to assume individuals choose beliefs in order to maximize their expected utility (given those chosen beliefs), subjects to a cost of choosing beliefs that are different from the true probabilities.

One particular way of implementing this idea is to suppose that given an objective belief $p$, the DM chooses a distorted belief $q$ which solves the following problem $$\max_{q\in\Delta(\Omega)}u\cdot q - \frac{1}{K}\Big[\sum_{\omega}q(\omega)[\ln q(\omega)-\Lambda\ln p(\omega) ]\Big],$$ where $p$ has full support.  

Here $K>0$, $\Lambda>0$ are parameters. The term $u\cdot q$ captures the anticipated expected utility of the DM, conditional on the distorted beliefs.  The term $\frac{1}{K}[\sum_{\omega}q(\omega)[\ln q(\omega) -\Lambda \ln p(\omega)]]$ is the cost of distorting beliefs to $q$, given objective beliefs $p$.  If $\Lambda=1$, then the cost is simply the Kullback-Liebler (KL) divergence between $p$ and $q$. In line with this explanation, we call the DM's problem a motivated beliefs problem with $\Lambda$-KL costs and utilities $u$ (this is essentially a subset of the set of preferences considered by \cite{bracha2012affective}, and when $\Lambda=1$ corresponds to preferences used in \cite{caplin2019wishful} and \cite{mayraz2019priors}).   

Notice that if instead the decision-maker instead faced the problem $\min_{q\in\Delta(\Omega)}u\cdot q + \frac{1}{K}[\sum_{\omega}q(\omega)[\ln(q(\omega))-\Lambda\ln(p(\omega))]]$, then this would be a subset of the ambiguity-averse variational preferences introduced by (\cite{maccheroni2006ambiguity}), and if $\Lambda=1$ this would be a multiplier preference (\cite{hansen2001robust}; \cite{strzalecki2011axiomatic}). In fact, our results would also apply in this situation.  

The next proposition summarizes there is an equivalence between the set of solutions to the motivated beliefs problem with $\Lambda$-KL costs, utilities $u$, and weighted power distortion functions (for related, recent, independent results, see \cite{dominiak2023inertial}; \cite{yang2023overprecise}; \cite{strzalecki2024variational}).   

\begin{proposition}\label{prop:mult}
The solution to a motivated beliefs problem as $p$ varies with $\Lambda$-KL costs and utilities $u$  is a weighted power distortion function.  Moreover, for a given weighted power distortion function $\phi$, there exists a utility function $u$ and two positive constants, $K$ and $\Lambda$, such that $\alpha = \Lambda$
and $\psi(\omega) = e^{Ku(\omega)}$.
\end{proposition}

Thus, when we use a generalized version of KL divergence as a cost function, and then have individuals optimize beliefs, the set of solutions is  equivalent to the set of weighted distortion  functions.  Just as in the previous subsection, this result links our distorted beliefs to several literatures on non-expected preferences, both in the world of motivated beliefs, and the world of ambiguity.


\subsection{Dynamic Consistency}

Our notion of coherency relates beliefs before and after the arrival of information.  Economists are often very interested in a related concept --- the consistency of preferences before and after receiving information --- what is typically termed \emph{dynamic consistency}.  A natural question to ask is what is the relationship between our coherent distorted beliefs and dynamic consistency.  

In order to explore this question we need to first relate beliefs to observed preferences.  Here, we will make the simplest possible assumption (as in Section \ref{sec:eu}): individuals act as expected utility maximizers conditional on their beliefs.  

Moreover, we need to fix a definition of dynamic consistency (for an early and influential survey, see \cite{machina1989dynamic}).  Informally, dynamic consistency says that the DMs preferences over future contingency plans agree with what he will do if those contingencies actually occur. There have been multiple ways in which dynamic consistency has been defined (e.g., compound lotteries as in \cite{karni1991atemporal,border1994dynamic,volij1994dynamic}} or in a Savage framework as in \cite{epstein1993dynamically}).  Recall that dynamic consistency specifically looks at preferences prior to the realization of events, and compares them to preferences after the realization of an event.  

Thus, in order to address dynamic consistency with respect to coherent beliefs, we will focus on the framework in Section \ref{sec:states}. Let us denote by $F$ the set of real-valued functions on $\Omega$.  Typical elements will be written $f$ and $g$.  
Given a distortion function $\phi:\Delta(\Omega)\rightarrow\Delta(\Omega)$ and belief $p\in\Omega$, define a function $V_{\phi}:F\times\Delta(\Omega)\rightarrow\mathbb{R}$ as follows: $V_\phi(f;p) = \sum_{\omega} f(\omega) \phi(p(\omega))$.  Consider an act $f_E g$, which pays off according to $f$ if a state in $E$ occurs, and according to $g$ if a state not in $E$ occurs.  In an abuse of notation let $p|E$ denote the probability distribution conditional on learning that event $E$ has occurred if Bayesian updating is used.  More generally, if we have a set of signals $\Theta$, let $p|\theta$ denote the probability distribution over states induced by the prior $p$ after learning signal $\theta$ and given Bayesian updating.

\begin{definition}
A distorted belief function $\phi$ induces Dynamically Consistent preferences if for all full support $p$ and $f,g,h$,  $V_\phi(f_E g;p) \geq V_\phi(h_E g;p)$ if and only if $V_\phi(f_E g;p|E) \geq V_\phi(h_E g;p|E)$.
  \end{definition}

Notice that if $\phi$ is the identity (that is, no distortion), then preferences are Dynamically Consistent.  The first comparison (the ex-ante preference) is simply $\sum_{\omega \in E} f(\omega) p(\omega) + \sum_{\omega \in E^C} g(\omega) p(\omega) \geq \sum_{\omega \in E} h(\omega) p(\omega) + \sum_{\omega \in E^C} g(\omega) p(\omega)$, while the second (the ex-post preference) is $\sum_{\omega \in E} f(\omega) \frac{p(\omega)}{\sum_{\omega'\in E}p(\omega')} \geq 
\sum_{\omega \in E} (\omega) h(\omega)\frac{p(\omega)}{\sum_{\omega'\in E}p(\omega')}$.  Moreover, it is clear that general distortion functions may not generate dynamically consistent behavior. 

Notice that, by the definition of coherency, conditioning on $E$ and then distorting is the same as distorting (to a set of beliefs) and then conditioning as a Bayesian would if the distortion function is power weighted.  Thus, it immediately follows that they induce Dynamically Consistent preferences.  The following proposition, stated without proof, claims that the converse is true as well.

\begin{proposition}
    Suppose $|\Omega| \geq 3$, $\phi$ is continuous and positive, and that preferences are represented by $V_{\phi}$ as discussed above. Then $\phi$ is a power-weighted distortion function if and only if preferences are dynamically consistent.
\end{proposition}

Similar exercises can be conducted using the concepts relating to experiments and signals.

\subsection{Idempotence, Fixed Points, and Iterates}\label{sec:limit}
We now turn to understanding what happens in the limit when the DM distorts their beliefs repeatedly.  As an initial result, we show, perhaps not surprisingly, that if $\phi$ is idempotent, then it must not involve any distortions at all---in other words $\phi(p) =p$.  We say that $\phi$ is \emph{idempotent} if $\phi \circ \phi = \phi$.

\begin{proposition}\label{prop:idempotent}Let $\phi$ be a power weighted distortion function. If $\phi$ is idempotent, then $\phi$ is the identity map. \end{proposition}

Of course, if $\phi$ is not idempotent, it raises the question of what the limit beliefs are after repeated distortions occur.  If we define $\phi^1=\phi$, and $\phi^n=\phi\circ \phi^{n-1}$, then $$\phi^n(p)(\omega)\propto_{\Omega} p(\omega)^{\alpha^n}\psi(\omega)^{\sum_{i=0}^{n-1}\alpha^i}.$$  

To explore limit beliefs, we can ask about the behavior of $\phi^n$ as $n$ gets large.  Clearly, $\phi$ is never a contraction (it maps point masses to point masses), but the iterative procedure still always converges to a fixed point.

\begin{proposition}\label{prop:limit}For any $\alpha>0$ and $\psi$, and all $p\in\Delta(\Omega)$, $\phi^*(p)=\lim_n \phi^n(p)$ exists, and has the following form:
\begin{enumerate}
\item If $0<\alpha<1$, then for all $\omega$ for which $p(\omega)>0$, $\phi^*(p)(\omega)\propto_{\Omega} \psi(\omega)^{\frac{1}{1-\alpha}}$.  Otherwise, $\phi^*(p)(\omega)=0$.
\item If $\alpha=1$, then for all $\omega$ for which $\psi(\omega)$ is maximal amongst the set of $\omega$ for which $p(\omega)>0$, $\phi^*(p)(\omega)\propto_{\Omega} p(\omega)$. Otherwise, $\phi^*(p)(\omega)=0$.
\item If $1 < \alpha$, then for all $\omega$ for which $p(\omega)^{\alpha-1}\psi(\omega)$ is maximal, we have $\phi^*(p)(\omega)\propto_{\Omega} \psi(\omega)^{\frac{1}{1-\alpha}}$.  Otherwise, $\phi^*(p)(\omega)=0$.
\end{enumerate}
\end{proposition}

Observe that the distorted beliefs $\phi^*$ characterized in Proposition~\ref{prop:limit} are coherent and idempotent, where since positivity may be violated, the coherent property only applies when both $p(E)>0$ and $\phi(p)(E)>0$.  Indeed, all $\phi^n$ are continuous, positive, and coherent distorted beliefs.  In the limit, the positivity and continuity properties can be lost, but idempotence is gained.  Of course, the only case positivity and continuity are not lost is, according to Proposition~\ref{prop:idempotent}, when $\phi^*$ is the identity map, which implies that $\phi$ itself is the identity.

When $\alpha \neq 1$, the probabilities of the form $p(\omega)\propto_{\Omega} \psi(\omega)^{\frac{1}{1-\alpha}}$ whenever $p(\omega)>0$ are exactly the fixed points of the functions $\phi$.  The behavior of the limit of $\phi^n$ is to always tend to one of these $2^{|\Omega|}-1$ fixed points:  for each $E\subseteq\Omega$, we have one such fixed point, which we can call $p^{\psi}_E$, where $p^{\psi}_E(\omega)\propto_{\Omega}\psi(\omega)^{\frac{1}{1-\alpha}}$ whenever $\omega\in E$, and otherwise $p^{\psi}_E(\omega)=0$.  

The difference is in the limiting behavior:  if $0<\alpha<1$, for example, all distributions with support $E$ will tend to $p^{\psi}_E(\omega)$.  We can call such a limiting rule a \emph{support rule}, as it is a constant rule as a function of the support of $p$.  

On the other hand when $1 < \alpha$, the \emph{only} distribution with support $E$ whose limit tends to $p^{\psi}_E$ is $p^{\psi}_E$ itself.  We can call such a limiting rule a \emph{maximum likelihood rule} as it is a constant rule as a function of the (weighted) maximum likelihood event, namely the set of $\omega$ for which $p(\omega)\psi^{\frac{1}{\alpha-1}}$.

Finally, when $\alpha =1$, the fixed points are of a different nature.  In this case, the $\psi$ function determines a weak order on states, whereby $\omega \succeq \omega'$ iff $\psi(\omega) \geq \psi(\omega')$.  For any $p$, the distorted belief $\phi^*(p)$ then proposes that we find the maximal states $\omega$ of this weak order amongst the set $p(\omega)>0$, and take the Bayesian update on the event consisting of this set of states.  The weak order can be thought of as partitioning $\Omega$ into events, which are then linearly ordered.  This is another kind of distorted belief, which in a sense is a Bayesian update on a fixed event, but specifies what happens when this fixed event has probability zero.  In this case, we move to a following event in the linear order and check whether it has nonzero probability, and so forth.  The fixed points of the original $\phi$ function are therefore all probability distributions whose support is contained completely in one of these events.  And we can call the limiting distorted belief 
a \emph{lexicographic coherent distorted belief}.

\bibliographystyle{plainnat}
\bibliography{reference}

\appendix

\section{Proofs}\label{app:proofs}

\noindent \textbf{Proof of Theorem \ref{thm:motivated}:}
We have shown that $\phi^{PW}$ satisfies all properties. We now show the other way. Let $\phi$ be a coherent distorted belief.  We will establish the result first on $\Delta_{++}(\Omega)$, the set of full support distributions.  Let us write the states as $\Omega = \{1,\ldots,n\}$.  Fix the three states, $1,2,n\in\Omega$.  We will define three maps $\phi_{1n}:\mathbb{R}_+\rightarrow\mathbb{R}_+$, and similarly $\phi_{12}$ and $\phi_{2n}$.  

For $q> 0$, define $\phi_{1n}(q)=\frac{\phi(p)(1)}{\phi(p)(n)}$ for any $p$ for which $\frac{p(1)}{p(n)}=q$.  Similarly define $\phi_{2n}(q)$ and $\phi_{12}(q)$.  Because $\phi$ is coherent, each of these maps is well-defined. To see this, suppose that $\frac{p(1)}{p(n)}=\frac{p'(1)}{p'(n)}$, where each of $p(1),p(n)>0$.  Then this means that $p(\cdot|\{1,n\})=p'(\cdot|\{1,n\})$.  Therefore $\phi(p)(\cdot|\{1,n\})=\phi(p(\cdot|\{1,n\}))=\phi(p'(\cdot|\{1,n\}))=\phi(p')(\cdot|\{1,n\})$.  Here, the first equality is Bayesian coherence, the second property is by assumption, and the third again by Bayesian coherence.  We may conclude that $\frac{\phi(p)(1)}{\phi(p)(n)}=\frac{\phi(p')(1)}{\phi(p')(n)}$, so that $\phi_{1n}$ is well-defined.

Now, observe that for any $q,r>0$, \begin{equation}\label{eq:pexider}\frac{\phi_{1n}(q)}{\phi_{2n}(r)}=\phi_{12}\left(\frac{q}{r}\right);\end{equation} this follows by considering any $p\in\Delta_{++}(\Omega)$ for which $\frac{p(1)}{p(n)}=q$ and $\frac{p(2)}{p(n)}=r$. Rewrite the equation~\eqref{eq:pexider} as $\phi_{1n}(ab)=\phi_{12}(a)\phi_{2n}(b)$, which is evidently a Pexider equation.  By Theorem 4 on p. 144 of \citet{aczel}, it has as solution:  $\phi_{1n}(x)=\pi(1) x^{\alpha}$ and $\phi_{2n}(y) = \pi(2) y^{\alpha}$. 
 In our case $\pi(1)>0$ and $\pi(2)>0$.  

Now, for any $m\notin \{1,2,n\}$, we can similarly establish an equation of the form: $\psi_{1n}(ab)=\phi_{1m}(a)\phi_{mn}(b)$, which also must have a solution of the form $\phi_{1n}(x)=\pi_m(1)x^{\alpha_m}$, $\phi_{mn}(y)=\pi(m) y^{\alpha_m}$.  But since $\phi_{1n}$ is the same equation in both cases, it must be that  $\pi_m(1)=\pi(1)$ and $\alpha_m=\alpha$.  Consequently, for all $i=\{1,\ldots,n-1\}$, we have $\phi_{in}(x)=\pi(i)x^{\alpha}$.  Now, for all $i\neq n$, define $\psi(i)=\frac{\pi(i)}{1+\sum_{j\neq n}\pi(j)}$ and define $\psi(n)=\frac{1}{1+\sum_{j\neq n}\pi(j)}$.  We have thus shown that for all $i\neq n$, we have $\frac{\phi(p)(i)}{\phi(p)(n)}=\frac{\psi(i)p(i)^{\alpha}}{\psi(n)p(n)^{\alpha}}$. Therefore, because probabilities sum to one, for any $p\in\Delta_{++}(\Omega)$, we have \[\phi(p)(\omega)=\frac{\psi(\omega)p(\omega)^{\alpha}}{\sum_{\omega'\in\Omega}\psi(\omega')p(\omega')^{\alpha}}.\] 

We now extend to all $\Delta(\Omega)$.  We claim that $\alpha > 0$.  To see why, let $\omega\in\Omega$, and let $p^n\in\Delta_{++}(\Omega)$ be any sequence for which $\lim_n p^n(\omega)=0$ and for all $\omega'\neq\omega$, $\lim_n p^n(\omega')=\frac{1}{\|\Omega\|}$.  Call this limit $p^*_{\omega}$.  Observe that, by positivity, $\phi(p^*_{\omega})(\omega)=0$.  But if $\alpha = 0$, $\lim_n \phi(p)(\omega)=\psi(\omega)>0$, and if $\alpha< 0$, $\lim_n \phi(p)(\omega) = 1$, in either case a contradiction.  So $\alpha > 0$ and the expression
\[\phi(p)(\omega)=\frac{\psi(\omega)p(\omega)^{\alpha}}{\sum_{\omega'\in\Omega}\psi(\omega')p(\omega')^{\alpha}}\]
holds for all $p\in\Delta(\Omega)$. $\Box$ 

\bigskip 

\noindent \textbf{Proof of Corollary \ref{cor:motivated}:} First let us show that for any $E\in\Pi$ and any $\omega,\omega'\in E$, $\psi(\omega)=\psi(\omega')$.  Because of our assumption (that there exists $E^*\in\Pi$ for which $1<|E^*|<|\Omega|$), there is $\omega^*\in\Omega\setminus E$.  Consider $p'=(1/2)\delta_{\omega}+(1/2)\delta_{\omega^*}$:  then $\phi(p)(\omega)=\frac{\psi(\omega)}{\psi(\omega)+\psi(\omega^*)}$.  Consider $p''=(1/2)\delta_{\omega'}+(1/2)\delta_{\omega^*}$.  Then $\phi(p')(\omega')=\frac{\psi(\omega')}{\psi(\omega')+\psi(\omega^*)}$.  By $\Pi$-marginality, and the fact that $\psi$ is positive, we then must conclude that $\psi(\omega)=\psi(\omega')$.

Now we show that $\alpha =1$.  Let $E^*$ be the event referenced in the hypothesis of the Corollary.  Let $\omega,\omega'\in E^*$ be distinct, and let $\omega^*\in\Omega\setminus E^*$.  Let $p'= (1/4)\delta_{\omega}+(1/4)\delta_{\omega'}+(1/2)\delta_{\omega^*}$ and let $p''=(1/2)\delta_{\omega}+(1/2)\delta_{\omega^*}$.  Then $\phi(p)(E^*)=\frac{\psi(\omega)2(1/4)^{\alpha}}{\psi(\omega)2(1/4)^{\alpha}+\psi(\omega^*)(1/2)^{\alpha}}$ and $\phi(p')(E^*)=\frac{\psi(\omega)(1/2)^{\alpha}}{\psi(\omega)(1/2)^{\alpha}+\psi(\omega^*)(1/2)^{\alpha}}$, conclude by $\Pi$-marginality that $2(1/4)^{\alpha}=(1/2)^{\alpha}$, or $2=2^{\alpha}$; that is, $\alpha = 1$. $\Box$

\bigskip

\noindent \textbf{Proof of Theorem~\ref{thm:grether}:} Let us first verify that equation~\eqref{eq:grether} is satisfied under the conditions.  Observe that for $p\in\Delta(\Omega)$ and experiment $\sigma$, $\mathcal{B}_{\sigma}(p,\theta)(\omega)=\frac{p(\omega)\sigma(\theta|\omega)}{\sum_{\omega'}p(\omega')\sigma(\theta|\omega')}$.  Therefore, $f(\mathcal{B}_{\sigma}(p,\theta))(\omega)=\frac{\psi(\omega) (p(\omega)\sigma(\theta|\omega))^{\alpha}}{\sum_{\omega'}\psi(\omega') (p(\omega')\sigma(\theta|\omega'))^{\alpha}}$.  On the other hand, $\mathcal{B}_{g \circ \sigma }(f(p),\theta)(\omega) = \frac{\psi(\omega)p(\omega)^{\alpha}\sigma(\theta|\omega)^{\alpha}}{\sum_{\omega'}\psi(\omega')p(\omega')^{\alpha}\sigma(\theta|\omega')^{\alpha}}$, evidently the same thing.

For the converse direction, first let $\theta\in\Theta$.  We claim that the quantity $g_{\omega}(\sigma(\cdot|\omega))(\theta)$ depends only on $\sigma(\theta|\omega)$.  To see why this is true, imagine two signal structures $\sigma$ and $\sigma'$ and let $\omega,\omega'\in\Omega$ for which $\omega\neq\omega'$, where $\sigma(\theta|\omega)=\sigma'(\theta|\omega')$.  Without loss, we may assume that $\sigma^*$ is a signal for which $\sigma^*(\cdot|\omega)=\sigma(\cdot|\omega)$ and $\sigma^*(\cdot|\omega')=\sigma'(\cdot|\omega')$.  For all remaining states $\omega^*$, assume for example that $\sigma^*(\theta|\omega^*)=\sigma(\theta|\omega)=\sigma'(\theta|\omega')$.  Then $\mathcal{B}_{\sigma^*}(p,\theta)$ is $p$, no matter what is $p$.  So $f(\mathcal{B}_{\sigma^*}(p,\theta))$ is simply $f(p)$, whereby we may conclude by equation~\eqref{eq:grether} that the map $\hat{\omega}\mapsto g_{\hat{\omega}}(\sigma^*(\cdot|\hat{\omega}))(\theta)$ is constant.  Then it is apparent that $g_{\omega}(\sigma(\cdot|\omega))(\theta)=g_{\omega}(\sigma^*(\cdot|\omega))(\theta)=g_{\omega'}(\sigma^*(\cdot|\omega'))(\theta)=g_{\omega'
}(\sigma'(\cdot|\omega'))(\theta)$.  Similarly, if $\omega=\omega'$ and and $\sigma(\theta|\omega)=\sigma(\theta|\omega')$, we may conclude that $g_{\omega}(\sigma(\cdot|\omega))(\theta)=g_{\omega'}(\sigma(\cdot|\omega))(\theta)$ by choosing any $\omega^*\neq \omega$ and a double application of the preceding argument.  So we have shown that $\sigma(\theta|\omega)=\sigma'(\theta|\omega')$ implies $g_{\omega}(\sigma(\cdot|\omega))(\theta)=g_{\omega'}(\sigma(\cdot|\omega')(\theta)$.

Thus, define $h_{\theta}:[0,1]\rightarrow \mathbb{R}$ via $h_{\theta}(p)=g_{\omega}(\sigma(\cdot|\omega))(\theta)$ for any $\omega\in\Omega$ and any $\sigma$ where $\sigma(\theta|\omega)=p$: the previous paragraph establishes that this map is well-defined.

Conclude that $$\mathcal{G}_{\sigma}(p,\theta)=\frac{f(p)(\omega)h_{\theta}(\sigma(\theta|\omega))}{\sum_{\omega'}f(p)(\omega')h_{\theta}(\sigma(\theta|\omega'))}$$ for some function $h_{\theta}:[0,1]\rightarrow [0,1]$. 

Next, let us show that there exists $h:[0,1]\rightarrow \mathbb{R}$ and for each $\theta\in\Theta$, $\gamma(\theta)>0$ for which $h_{\theta}(q)=\gamma(\theta)h(q)$.  By Lemma 2.1 in \cite{jamison1976}, there is some $p^*\in \Delta(\Omega)$ for which $f(p^*)$ is a uniform distribution over $\Omega$.  Thus for any $\sigma$ and any $\theta\in\Theta$, $\mathcal{G}_{\sigma}(p^*,\theta)=\frac{h_{\theta}(\sigma(\theta|\omega))}{\sum_{\omega'}h_{\theta}(\sigma(\theta|\omega'))}$.  Now, for any signal $\sigma$, define a new signal $\sigma_{\theta_1,\theta_2}$ by $\sigma_{\theta_1,\theta_2}(\theta_1|\omega)=\sigma(\theta_2|\omega)$ and $\sigma_{\theta_1,\theta_2}(\theta_2|\omega)=\sigma(\theta_1|\omega)$, where finally $\sigma_{\theta_1,\theta_2}(\theta|\omega)=\sigma(\theta|\omega)$ for all $\theta\in\Theta\setminus\{\theta_1,\theta_2\}$.  Then clearly $\mathcal{B}_{\sigma}(p^*,\theta_1)=\mathcal{B}_{\sigma_{\theta_1,\theta_2}}(p^*,\theta_2)$.  Consequently by equation~\eqref{eq:grether}, \begin{equation}\label{eq:grether2}\frac{h_{\theta_1}(\sigma(\theta_1|\omega))}{\sum_{\omega'}h_{\theta_1}(\sigma(\theta_1|\omega'))}=\frac{h_{\theta_2}(\sigma_{\theta_1,\theta_2}(\theta_2|\omega))}{\sum_{\omega'}h_{\theta_2}(\sigma_{\theta_1,\theta_2}(\theta_2|\omega'))}=\frac{h_{\theta_2}(\sigma(\theta_1|\omega))}{\sum_{\omega'}h_{\theta_2}(\sigma(\theta_1|\omega'))},\end{equation} where the second equality is by definition of $\sigma_{\theta_1,\theta_2}$.

So, for any $q\in(0,1]$, choose $\sigma$ and $\omega_1,\omega_2$ for which $\sigma(\theta_1|\omega_1)=q$, $\sigma(\theta_1|\omega_2)=1$ and $\sigma(\theta_1|\omega)=0$ for all $\omega\in\Omega\setminus\{\omega_1,\omega_2\}$, which can be done as $|\Theta|\geq 2$.  We conclude by equation~\eqref{eq:grether2} that $$\frac{h_{\theta_1}(q)}{h_{\theta_2}(q)}=\frac{h_{\theta_1}(q)+h_{\theta_1}(1)}{h_{\theta_2}(q)+h_{\theta_2}(1)}=\frac{h_{\theta_1}(1)}{h_{\theta_2}(1)}.$$  Consequently for any pair $\theta_1,\theta_2$, we have $h_{\theta_1}(q)=\frac{h_{\theta_1}(1)}{h_{\theta_2}(1)}h_{\theta_2}(q)$, which establishes (by fixing some $\theta^*\in\Theta$ and defining $h=h_{\theta^*}$) that there is positive and continuous $h:[0,1]\rightarrow\mathbb{R}$ and $\gamma(\theta)>0$ such that for all $\theta\in\Theta$, $h_{\theta}(q)=\gamma(\theta)h(q)$.

Equation~\eqref{eq:grether} now reads $$f\left(\frac{p(\cdot)\sigma(\theta|\cdot)}{\sum_{\omega'}p(\omega')\sigma(\theta|\omega')}\right)(\omega)=\frac{f(p)(\omega)h(\sigma(\theta|\omega))}{\sum_{\omega'}f(p)(\omega')h(\sigma(\theta|\omega'))},$$ as $\gamma(\theta)$ factors out. Taking the uniform prior and denoting its output by $\psi$, we then find that (assuming all relevant variables are nonzero) for any $\omega,\omega^*$,  $$\frac{f\left(\frac{\sigma(\theta|\cdot)}{\sum_{\omega'}\sigma(\theta|\omega')}\right)(\omega)}{f\left(\frac{\sigma(\theta|\cdot)}{\sum_{\omega'}\sigma(\theta|\omega')}\right)(\omega^*)}=\frac{\psi(\omega)h(\sigma(\theta|\omega))}{\psi(\omega^*)h(\sigma(\theta|\omega^*))}.$$

Clearly this implies that \begin{equation}\label{eq:fequation}f(p)(\omega) = \frac{\psi(\omega)h(\beta p(\omega))}{\sum_{\omega'}\psi(\omega')h(\beta p(\omega'))}\end{equation} for any $\beta \leq 1$, by choosing $\sigma$ and $\theta\in\Theta$ for which $\sigma(\theta|\omega)=\beta p(\omega)$.  Considering the case of $\beta^* = 1$ and an arbitrary $\beta$, we obtain \begin{equation}\label{eq:grether3}\frac{h(p(\omega))}{h(\beta p(\omega))}=\frac{\sum_{\omega'}\psi(\omega')h(p(\omega'))}{\sum_{\omega'}\psi(\omega')h(\beta p(\omega'))}.\end{equation}

Now, for any $q<1$ and $r \leq 1- q$, we may choose $p\in\Delta(\Omega)$ and $\omega_1,\omega_2$ for which $p(\omega_1)=q$ and $p(\omega_2)=r$, since $|\Omega|\geq 3$.  Consequently equation~\eqref{eq:grether3} implies that $\frac{h(q)}{h(\beta q)}=\frac{h(r)}{h(\beta r)}$.

So, if $q',q<1$, we obtain by choosing $r \leq \min\{1-q,1-q'\}$, that $\frac{h(q)}{h(\beta q)}=\frac{h(r)}{ h(\beta r)}=\frac{h(q')}{ h(\beta q')}$. By appealing to continuity, we know that for any $q,q'\in (0,1]$ and $\beta \in (0,1]$, we have $\frac{h(q)}{h(\beta q)}=\frac{h(q')}{h(\beta q')}$.  By choosing $q' = 1$, we obtain $h(\beta q)=\frac{h(\beta)h(q) }{h(1)}$, for any $\beta,q\in (0,1]$.  

Standard techniques establish that this equation has a solution given by $h(q) = h(1)q^{\alpha}$, where $\alpha \neq 0$.\footnote{That is, we can rewrite $m(x) = \frac{h(x)}{h(1)}$ and observe that $m(xy) = m(x)m(y)$ for all $x,y\in (0,1]$.  For any $x > 1$, define $m(x) = (m(1/x))^{-1}$ and observe that the equation $m(xy)=m(x)m(y)$ holds for all $x,y>0$, while remaining continuous.  Then apply Theorem 2 of p. 41 of \cite{aczel} to obtain that $m(x) = x^{\alpha}$, so that $h(x)=h(1)x^{\alpha}$.}

By suitably renormalizing $h$ and the coefficients $\gamma(\theta)$, we may assume without loss that $h(q) = q^{\alpha}$.  In order that continuity at $0$ remain satisfied, we clearly must have $\alpha > 0$.  We conclude by observing $f$ has the appropriate form by equation~\eqref{eq:fequation} as applied with $\beta =1$, and $g_{\omega}$ has the appropriate form, by the definition of $h_{\theta}$. $\Box$

\bigskip

\noindent \textbf{Proof of Corollary \ref{cor:grether2}:} The constraint $\sum_{\theta}g_{\omega}(\sigma(\theta|\omega))=1$ implies that $\gamma(\theta)=1$ for each $\theta$, by considering any $\sigma$ for which $\sigma(\theta|\omega)=1$.  Now, by choosing any pair $\theta,\theta'$ and $\sigma$ for which $\sigma(\theta'|\omega)=\sigma(\theta|\omega)=1/2$, we get $(1/2)^{\alpha}+(1/2)^{\alpha}=1$, which clearly implies $\alpha =1$.  The rest follows immediately from Theorem~\ref{thm:grether}.$\Box$

\bigskip 

\noindent \textbf{Proof of Theorem \ref{thm:verygeneral}:}
Notationally, in this proof we will write $p(\cdot;\theta)$ to refer to the restriction of $p(\cdot|\theta)$ to $\Omega\times\{\theta\}$.  Generally then, $p(\cdot;\theta)\in\Delta(\Omega)$ whereas $p(\cdot|\theta)\in\Delta(\Omega\times\Theta)$.  Similarly for $p(\cdot;\omega)$.  Observe that, by the rules of probability, for any $p\in\Delta(\Omega\times\Theta)$ for which $p(\theta)>0$, $\phi(p)(\omega,\theta)=\phi(p)(\theta)\phi(p)(\omega;\theta)$.

We claim that for each $\theta\in\Theta$, there is $h_{\theta}:\Delta(\Omega)\rightarrow\Delta(\Omega)$ which is both positive and continuous for which $\phi(p)(\omega;\theta)=h_{\theta}(p(\cdot;\theta))$.  Suppose $p,p'\in\Delta(\Omega\times\Theta)$ satisfy $p(\cdot;\theta)=p'(\cdot;\theta)$, where $p(\theta),p'(\theta)>0$.  Then $\phi(p)(\cdot|\theta)=\phi(p(\cdot|\theta))=\phi(p'(\cdot|\theta))=\phi(p')(\cdot|\theta)$, where the first and third equalities are by weak coherence.   Consequently we may write $$\phi(p)(\omega,\theta)=\phi(p)(\theta)h_{\theta}(p(\omega;\theta)).$$


Fix any $q\in \Delta(\Omega)$ and let $\theta^*\in\Theta$.  Define $p(\omega,\theta^*)=q(\omega)$ for all $\omega$, and for any $\theta\neq\theta^*$, $p(\omega,\theta)=0$.  Then owing to marginality, $\phi^M(q)(\omega)=\phi(p)(\omega)=h_{\theta^*}(p(\cdot;\theta))(\omega)=h_{\theta^*}(q)(\omega)$, where the second equality follows by positivity.  Since $\theta^*$ and $q$ were arbitrary, we may conclude that $h_{\theta}=\phi^M$ for all $\theta\in\Theta$.


So, we infer that \begin{equation}\label{eq:thisequation}\phi(p)(\omega,\theta)=\phi(p)(\theta)\phi^M(p(\omega;\theta))\end{equation} and further that $\phi(p)(\cdot;\theta)=\phi^M(p(\cdot;\theta))$.

Let $\varnothing\neq A\subseteq \Omega$.  Recall that $|\Theta|\geq 2$.  We claim that for any $q\in\Delta(\Omega)$, we have that $\phi^M(q(\cdot|A))=\phi^M(q)(\cdot|A)$.  This follows by fixing any $\theta,\theta'\in \Theta$ for which $\theta\neq\theta'$ and a pair $p,p'\in\Delta(\Omega\times\Theta)$ for which 
$p$ is defined so that: $p(\omega,\theta)=q(\omega)$ (all $\theta^*\neq \theta$ have $p(\theta^*)=0$) and $p'$ defined as:  for any $\omega\in A$, $p'(\omega,\theta)=q(\omega)$, and for any $\omega\notin A$, $p'(\omega,\theta')=q(\omega)$, all remaining $(\omega^*,\theta^*)$ assigned probability zero.  Observe that for each $\omega\in\Omega$, $p(\omega)=p'(\omega)$. 
By positivity of $\phi$, for any $\omega\in\Omega$, equation~\eqref{eq:thisequation} implies \begin{equation}\label{eq:thatequation}\phi(p')(\omega)=\phi(p')(\theta)\phi^M(q(\cdot|A))(\omega)+\phi(p')(\theta')\phi^M(q(\cdot|\Omega\setminus A))(\omega).\end{equation}   By positivity of $\phi^M$, we must have that $$\phi^M(q)=\phi(p)(\cdot;\theta)=\phi(p)(\theta)\phi(p)(\cdot;\theta)=\sum_{\theta^*}\phi(p)(\theta^*)\phi(p)(\cdot;\theta^*)=\phi(p)|_{\Omega}.$$ Here, the first equality is by definition, the second because $p(\theta)=1$ implies $\phi(p)(\theta)=1$ by positivity, the third because for all $\theta^*\neq \theta$, $p(\theta^*)=0$ implies $\phi(p)(\theta^*)=0$ by positivity, and the last by the rules of probability.  Further, by marginality, $\phi(p)|_{\Omega}=\phi(p')|_{\Omega}$.  Finally, $\phi(p')|_{\Omega}=\phi(p')(\theta)\phi^M(q(\cdot|A))+\phi(p')(\theta')\phi^M(q(\cdot|\Omega\setminus A))$ by equation~\eqref{eq:thatequation}.  Conclude that for any $\omega$, $\phi^M(q)(\omega)=\phi(p')(\theta)\phi^M(q(\cdot|A))(\omega)+\phi(p')(\theta')\phi^M(q(\cdot|\Omega\setminus A))(\omega) $.  This in particular implies by positivity of $h$ that if $\omega\in A$, $\phi^M(q)(\omega)=\phi(p')(\theta)\phi^M(q(\cdot|A))(\omega)$.  This obviously implies that $\phi^M(q)(\cdot|A)=\phi^M(q(\cdot|A))$.

Applying Theorem~\ref{thm:motivated}, we can conclude that $\phi^M(q)(\omega)\propto_{\Omega} \psi(\omega)q(\omega)^{\alpha}$ for some $\alpha > 0$ and for all $\omega\in\Omega$, some $\psi(\omega)>0$. 


We claim that $\alpha = 1$, without loss, assume that there are exactly three states, $\{\omega_1,\omega_2,\omega_3\}$ (otherwise, we assign zero probability to the remaining states).  Fix two signals $\{\theta_1,\theta_2\}$.  We may without loss assume that $\psi(\omega_1)+\psi(\omega_2)+\psi(\omega_3)=1$.  
Now consider $p\in \Delta(\Omega\times\Theta)$ where $p(\omega_1,\theta_1)=1/3$, $p(\omega_2,\theta_1)=1/6$, $p(\omega_3,\theta_1)=0$, $p(\omega_1,\theta_2)=0$, $p(\omega_2,\theta_2)=1/6$, $p(\omega_3,\theta_2)=1/3$, where all unspecified pairs have probability $0$. 

Observe that $(p(\omega_1|\theta_1),p(\omega_2|\theta_1),p(\omega_3|\theta_1))=(2/3,1/3,0)$, which we can succinctly write as $p(\cdot|\theta_1)=(2/3,1/3,0)$.  Similarly, $p(\cdot|\theta_2)=(0,1/3,2/3)$.  Finally $(p(\omega_1),p(\omega_2),p(\omega_3))=(1/3,1/3,1/3)$.

Now, simple calculation establishes that $\phi(p)(\omega_1)=\frac{\phi(p)(\theta_1)\psi(\omega_1)(2/3)^{\alpha}}{\psi(\omega_1)(2/3)^{\alpha}+\psi(2)(1/3)^{\alpha}}$; this uses Bayes rule in the form $\phi(p)(\omega_1)=\sum_{\theta\in\Theta}\phi(p)(\theta)\phi(p)(\omega_1|\theta)=\sum_{\theta\in\Theta}\phi(p)(\theta)h(p(\cdot;\theta)
)(\omega_1)=\phi(p)(\theta_1)h(p(\cdot;\theta_1))(\omega_1)$, where the last equality follows by positivity.  Similarly, $h(p|_{\Omega})(\omega_1)=\psi(\omega_1)$ (owing to the fact that $\psi(\omega_1)+\psi(\omega_2)+\psi(\omega_3)=1$).  As previously observed, marginality implies that $\phi(p)(\omega_1)=h(p|_{\Omega})(\omega_1)$, which after simplifying the preceding expressions, establishes:\begin{equation}\label{eq:alphaone}\phi(p)(\theta_1)2^{\alpha}=\psi(\omega_1)2^{\alpha}+\psi(\omega_2).\end{equation}

Symmetrically, marginality implies that $\phi(p)(\omega_3)=h(p|_{\Omega})(\omega_3)$ and simplifying, we obtain:\begin{equation}\label{eq:alphatwo}\phi(p)(\theta_2)2^{\alpha}=\psi(\omega_3)2^{\alpha}+\psi(\omega_2).\end{equation} 

Now sum equations~\eqref{eq:alphaone} and \eqref{eq:alphatwo}, using the fact that $\phi(p)(\theta_1)+\phi(p)(\theta_2)=1$ and $\psi(\omega_1)+\psi(\omega_2)+\psi(\omega_3)=1$ to obtain: $$2^{\alpha}=(1-\psi(\omega_2))2^{\alpha}+2\psi(\omega_2).$$  

After cancelling terms it becomes apparent as $\psi(\omega_2)>0$ that we must have $2^{\alpha}=2$, or $\alpha=1$.

So we have concluded that $\phi^M(q)(\omega)\propto_{\Omega} \psi(\omega)q(\omega)$, which establishes the first part of the result.  It remains to show that $\phi(p)(\theta)=p(\theta)$ under the additional hypotheses.  

Suppose first that $|\Omega|\geq |\Theta|$.  Let us consider any $p$ for which for all $\theta\in\Theta$, $p(\theta)>0$ and for which the set $\{p(\cdot;\theta)\}_{\theta\in\Theta}$ is linearly independent.  Owing to equation~\eqref{eq:thisequation}, it follows that \begin{equation}\label{eq:representation2}\phi(p)(\omega,\theta)\propto_{\Omega\times\Theta} \phi(p)(\theta)\psi(\omega)p(\omega|\theta)=\frac{\phi(p)(\theta)\psi(\omega)p(\omega|\theta)}{\sum_{\omega',\theta'}\phi(p)(\theta')\psi(\omega')p(\omega'|\theta')}.\end{equation}

By marginality, then we know that for every $\omega\in\Omega$, $$\frac{\sum_{\theta'}\phi(p)(\theta')\psi(\omega)(\theta')p(\omega|\theta')}{\sum_{\omega',\theta'}\phi(p)(\theta')\psi(\omega')p(\omega'|\theta')}=\frac{\sum_{\theta'}p(\theta')\psi(\omega)p(\omega|\theta')}{\sum_{\omega',\theta'}p(\theta')\psi(\omega')p(\omega'|\theta')}.$$  This follows by considering any $p'\in\Delta(\Omega\times\Theta)$ for which for some $\theta\in\Theta$, $p'(\omega,\theta)=\sum_{\theta'}p(\omega,\theta')=\sum_{\theta'}p(\theta')p(\omega|\theta')$ and observing that marginality implies $\phi(p)(\omega)=\phi(p')(\omega)$, whence we use the formula for $\phi$. 

This demonstrates that \begin{equation}\label{eq:linearindependent}\sum_{\theta'}\phi(p)(\theta')\psi(\omega)p(\omega|\theta')\propto_{\Omega} \sum_{\theta'}p(\theta')\psi(\omega)p(\omega|\theta').\end{equation}  Denote by $p^{\psi}_{\theta}$ the vector in $\mathbb{R}^{\Omega}$ (not a probability measure) for which $p^{\psi}_{\theta}(\omega)=p(\omega|\theta)\psi(\omega)$.  Now, because the set of vectors $\{p(\cdot;\theta)\}_{\theta\in\Theta}$ is linearly independent in $\mathbb{R}^{\Omega}$, so is the set of vectors $\{p^{\psi}_{\theta}\}_{\theta\in\Theta}$.  Because of Equation~\eqref{eq:linearindependent} and linear independence, we may therefore conclude that there is some $\lambda > 0$ for which $\phi(p)(\theta)=\lambda p(\theta)$.  Because $\sum_{\theta}\phi(p)(\theta)=1$, we may conclude $\phi(p)(\theta)=p(\theta)$.

Now, for arbitrary $p\in\Delta(\Omega\times\Theta)$, the result follows by approximation and continuity of $\phi$:  because $|\Omega|\geq |\Theta|$, any $p\in\Delta(\Omega\times\Theta)$ can be approximated by $p'\in\Delta(\Omega\times\Theta)$ for which for all $\theta\in\Theta$, $p'(\theta)>0$ and $\{p'(\cdot;\theta)\}_{\theta\in\Theta}$ is linearly independent, demonstrating that $\phi(p)(\theta)=p(\theta)$.

Consider finally the case in which $|\Theta|\geq 3$ and $\phi$ is strongly signal coherent.  Then for each $\theta\in\Theta$, let $p_{\theta}\in\Delta(\Omega)$ for which for all $\theta,\theta'\in\Theta$, $p_{\theta}\neq p_{\theta'}$.  Define $f^*:\Delta(\Theta)\rightarrow\Delta(\Theta)$ in the following way.  For each $q\in\Delta(\theta)$, define $p_q(\omega,\theta)=q(\theta)p_{\theta}(\omega)$.  Then define $f^*(q)(\theta)=\phi(p_q)(\theta)$.  Strong signal coherence implies that $f^*$ satisfies the hypotheses of Theorem~\ref{thm:motivated}.  Consequently we can say that there is $\alpha>0$ and for each $\theta\in\Theta$, some $\xi(\theta)>0$ so that $f^*(q)(\theta)\propto_{\Theta} \xi(\theta)(q(\theta))^{\alpha}$.  Thus, according to our earlier notation, $$\phi(p_q)(\theta)\propto_{\Theta} \xi(\theta)(p_q(\theta))^{\alpha}=\xi(\theta)(q(\theta))^{\alpha}.$$

Now, let $\theta,\theta'\in\Theta$ with $\theta\neq\theta'$.  Let $q\in\Delta(\Theta)$ for which $q(\{\theta,\theta'\})=1$.  Marginality implies that, using the representation in Equation~\eqref{eq:representation2} and as usual choosing some $p'$ concentrating support on some fixed $\theta^*\in\Theta$ with $p|_{\Omega}=p'|_{\Omega}$, $$\psi(\omega)[\xi(\theta)(q(\theta))^{\alpha}p_{\theta}(\omega)+\xi(\theta')(q(\theta'))^{\alpha}p_{\theta'}(\omega)]\propto_{\Omega} \psi(\omega)[q(\theta)p_{\theta}(\omega)+q(\theta')p_{\theta'}(\omega)]$$ which implies, by cancelling out the $\psi$ terms, \begin{equation}\label{eq:newprop}\xi(\theta)(q(\theta))^{\alpha}p_{\theta}(\omega)+\xi(\theta')(q(\theta'))^{\alpha}p_{\theta'}(\omega)\propto_{\Omega}q(\theta)p_{\theta}(\omega)+q(\theta')p_{\theta'}(\omega).\end{equation}

Equation~\eqref{eq:newprop} and the fact that $p_{\theta}\neq p_{\theta'}$ jointly imply that for any such $(q(\theta),q(\theta'))$ pair, there is $\lambda_q\in\mathbb{R}_{++}$ for which $$(\xi(\theta)(q(\theta))^{\alpha},\xi(\theta')(q(\theta'))^{\alpha})=\lambda_q (q(\theta),q(\theta')).$$ In other words $$\frac{\xi(\theta)(q(\theta))^{\alpha}}{\xi(\theta')(q(\theta'))^{\alpha}}=\frac{q(\theta)}{q(\theta')}.$$  This expression holds no matter which $q$ satisfying $q(\{\theta,\theta'\})=1$ we choose, which implies that $\alpha = 1$ and $\xi(\theta)=\xi(\theta')$.  In fact, since $\theta,\theta'$ were arbitrary, it follows that $\xi$ is constant as a function of $\theta$.

Therefore, we may conclude that if $q(\theta)>0$ and $q(\theta')>0$, then $\frac{\phi(p_q)(\theta)}{\phi(p_q)(\theta')}=\frac{q(\theta)}{q(\theta')}$.  Obviously this implies that for any $q\in\Delta(\Theta)$, $\phi(p_q)(\theta)=q(\theta)$.  

Now, we had initially chosen $\{p_{\theta}\}_{\theta\in\Theta}$ to satisfy the hypothesis that $p_{\theta}\neq p_{\theta'}$ whenever $\theta\neq\theta'$.  So in fact, for any $p\in\Delta(\Omega\times\Theta)$ for which $p|_{\theta}\neq p'_{\theta'}$ when $\theta\neq \theta'$, we have shown that $\phi(p)(\theta)=p(\theta)$.  The general result now follows by continuity as any $p\in\Delta(\Omega\times\Theta)$ is arbitrarily close to $p'\in\Delta(\Omega\times\Theta)$ for which $p'_{\theta}\neq p'_{\theta'}$ when $\theta\neq\theta'$.  $\Box$

\bigskip





\noindent \textbf{Proof of Theorem \ref{thm:weakcont}:} We apply Theorem~\ref{thm:motivated}.  Let us fix any $X'\subseteq X$ for which $3\leq |X|<+\infty$ and $\underline{x}\in X'$.

By positivity, we can define an induced map $\phi_{X'}:\Delta(X')\rightarrow \Delta(X')$ in the natural way, as for any $p\in\Delta(X')$, there exists a unique $p^*\in\Delta(X)$ for which for all $x\in X'$, $p(x)=p^*(x')$.  Positivity tells us that the support of $\phi(p^*)$ is a subset of $X'$, so that we may define $\phi_{X'}(p)=\phi(p^*)|_{X'}$.

It is simple to verify that the axioms of Theorem~\ref{thm:motivated} are satisfied for $\phi_{X'}$, so that there exists $\psi_{X'}:X'\rightarrow\mathbb{R}_{++}$ and $\alpha_{X'}>0$ as described in that result.  Without loss, we may assume that $\psi_{X'}(\underline{x})=1$ for each $X'$.  It is standard to establish that, with this restriction, the representation afforded by Theorem~\ref{thm:motivated} is unique.

Further, it is straightforward to establish that for any $X'$ and $X''$ finite with at least three elements, $\alpha_{X'}=\alpha_{X''}$.  This follows by considering $X' \cup X''$, with associated $\alpha_{X'\cup X''}$ and appealing to the uniqueness result described in the previous paragraph, to obtain $\alpha_{X'}=\alpha_{X'\cup X''}=\alpha_{X''}$.  Similarly, we may define $\psi:X\rightarrow\mathbb{R}$ via $\psi(x)=\psi_{X'}(x)$ for any $X'$ finite with at least three elements, for which $x\in X'$.  Again the uniqueness result establishes that $\psi$ is well-defined.

So, let $\alpha = \alpha_{X'}$ for any $X'$ finite with at least three elements. Finally let $p\in\Delta(X)$ be arbitrary, and let $X'$ contain the support of $p$ (and at least three elements), we then obtain that for any $x'$ in the support of $p$:  $\phi(p)(x') = \frac{\psi(x')p(x')}{\sum_{x\in X'}\psi(x)p(x)}$.  Observe then that for any $x'$ in the support of $p$, $\phi(p)(x')=\frac{\psi(x')p(x')}{\sum_{x:p(x)>0}\psi(x)p(x)}$.  Similarly, by positivity, if $x'$ is not in the support of $p$, $0=\phi(p)(x')=\phi(p)(x')=\frac{\psi(x')p(x')}{\sum_{x:p(x)>0}\psi(x)p(x)}$.

Next, we establish that $\psi$ is continuous.  Let $x_n \in X$, $x_n \rightarrow x^*$.  We want to claim that $\psi(x_n) \rightarrow \psi(x^*)$.  To this end, fix $x'\in X\setminus \{x^*,x_1,x_2\ldots\}$ and let $p_n$ be a lottery placing probability $.5$ on $x_n$ and $.5$ on $x'$; while $p^*$ places probability $.5$ on $x^*$ and $.5$ on $x'$.  Observe that $p_n \xrightarrow{\textit{ w }}p^*$.  Consequently, $\phi(p_n)\xrightarrow{\textit{ w }}\phi(p^*)$ by weak continuity.  In particular it is clear that $\phi(p_n)(x_n)\rightarrow \phi(p^*)(x^*)$, by the definition of weak convergence.  Therefore: $\frac{\psi(x_n)}{\psi(x_n)+\psi(x')}\rightarrow \frac{\psi(x^*)}{\psi(x^*)+\psi(x')}$, which clearly implies $\psi(x_n)\rightarrow\psi(x^*)$, verifying continuity of $\psi$.

Finally, we establish that $\alpha =1$.  Let $x_n \rightarrow x^*$, and let $x'\in X\setminus \{x^*,x_1,x_2,\ldots\}$.  Let $q_n$ place probability $1/3$ on $x'$, $1/3$ on $x_n$ and $1/3$ on $x^*$, and let $q^*$ place probability $1/3$ on $x'$ and $2/3$ on $x^*$.  Clearly, $q_n \xrightarrow{\textit{ w }}q^*$.  So by weak continuity, $\phi(q_n)\rightarrow \phi(q^*)$.  In particular, by the definition of weak convergence, this implies that $\frac{(1/3)^{\alpha}[\psi(x_n)+\psi(x^*)]}{(1/3)^{\alpha}[\psi(x_n)+\psi(x^*)+\psi(x')]}\rightarrow \frac{(2/3)^{\alpha}\psi(x^*)}{(2/3)^{\alpha}\psi(x^*)+(1/3)^{\alpha}\psi(x')}$.  Now, owing to the fact that $\psi$ is continuous, we also know that $\frac{(1/3)^{\alpha}[\psi(x_n)+\psi(x^*)]}{(1/3)^{\alpha}[\psi(x_n)+\psi(x^*)+\psi(x')]}\rightarrow \frac{2(1/3)^{\alpha}\psi(x^*)}{2(1/3)^{\alpha}\psi(x^*)+(1/3)^{\alpha}\psi(x')}$.  Therefore, $\frac{2(1/3)^{\alpha}\psi(x^*)}{2(1/3)^{\alpha}\psi(x^*)+(1/3)^{\alpha}\psi(x')}=\frac{(2/3)^{\alpha}\psi(x^*)}{(2/3)^{\alpha}\psi(x^*)+(1/3)^{\alpha}\psi(x')}$.  Simple algebra then verifies that $\alpha =1$.

The converse is standard and is omitted. $\Box$

\bigskip

\noindent \textbf{Proof of Proposition \ref{prop:mult}:}
The first order condition is that for each $\omega\in\Omega$, $$u(\omega) +\rho -\frac{1}{K}[\ln(q(\omega))-\Lambda\ln(p(\omega))]=0.$$  Here $\rho$ is a Lagrange multiplier which also incorporates the term from differentiating $\ln(q(\omega))$.

We then have that for all $\omega,\omega'$, $$Ku(\omega)-Ku(\omega')=\ln\left(\frac{q(\omega)}{p(\omega)^{\Lambda}}\frac{p(\omega')^{\Lambda}}{q(\omega')}\right).$$

So $$\frac{e^{Ku(\omega)}}{e^{Ku(\omega')}}=\frac{q(\omega)}{p(\omega)^{\Lambda}}\frac{p(\omega')^{\Lambda}}{q(\omega')}.$$  Rewriting: $$\frac{e^{Ku(\omega)}p(\omega)^{\Lambda}}{e^{Ku(\omega')}p(\omega')^{\Lambda}}=\frac{q(\omega)}{q(\omega')},$$ which is of the form in Theorem~\ref{thm:motivated}.

Observe that since $\xi(q)=q\ln(q)$ is a convex function, it follows that for any $\Lambda > 0$ and any $p\in\Delta_{++}(\Omega)$, $\sum_{\omega}q(\omega)[\ln(q(\omega))-\Lambda\ln(p(\omega))]$ is also convex, which ensures the underlying optimization problem is a concave optimization problem.$\Box$

\bigskip

\noindent \textbf{Proof of Proposition  \ref{prop:idempotent}:} We will show that $\psi(\omega)=\psi(\omega')$ for all $\omega,\omega'$ and that $\alpha =1$.  To this end, idempotence implies that for all $\omega,\omega'$ and all $p\in\Delta(\Omega)$, if $p(\omega),p(\omega')>0$, then $$\frac{\psi(\omega)p(\omega)^{\alpha}}{\psi(\omega')p(\omega')^{\alpha}}=\frac{\psi(\omega)^{\alpha+1}p(\omega)^{\alpha^2}}{\psi(\omega')^{\alpha+1}p(\omega')^{\alpha^2}}.$$  Thus, $$\left(\frac{\psi(\omega)}{\psi(\omega')}\right)^{\alpha}\left(\frac{p(\omega)}{p(\omega')}\right)^{\alpha^2-\alpha}=1.$$  Since this holds for each $p$, the left hand side is constant as a function of $p$, which can only be if $\alpha^2-\alpha = 0$; since $\alpha > 0$, we obtain $\alpha=1$.  Further it then follows that $\psi(\omega)=\psi(\omega')$. $\Box$

\bigskip

\noindent \textbf{Proof of Proposition \ref{prop:limit}:} Recall $\phi^n(p)(\omega)\propto_{\Omega} p(\omega)^{\alpha^n}\psi(\omega)^{\sum_{i=0}^{n-1}\alpha^i}$.  That is, for any $\omega\in\Omega$ for which $p(\omega)>0$, we have $$\phi^n(p)(\omega)=\frac{p(\omega)^{\alpha^n}\psi(\omega)^{\sum_{i=0}^{n-1}\alpha^i}}{\sum_{\omega'}p(\omega')^{\alpha^n}\psi(\omega')^{\sum_{i=0}^{n-1}\alpha^i}}.$$

If $0 < \alpha < 1$, then because $\lim_n \sum_{i=0}^{n-1}\alpha^i=\frac{1}{1-\alpha}$, and $\lim_n \alpha^n \rightarrow 0$, the formula directly follows.

If $1 \leq \alpha$, rewrite  $$\phi^n(p)(\omega)\propto_{\Omega} [p(\omega)^{\alpha-1}\psi(\omega)]^{\sum_{i=0}^{n-1}\alpha^i}p(\omega).$$  In other words, for all $\omega$ for which $p(\omega)>0$, we have $$\phi^n(p)(\omega)=\frac{[p(\omega)^{\alpha-1}\psi(\omega)]^{\sum_{i=0}^{n-1}\alpha^i}p(\omega)}{\sum_{\omega'}[p(\omega')^{\alpha-1}\psi(\omega')]^{\sum_{i=0}^{n-1}\alpha^i}p(\omega')}.$$

Observe that when $\alpha=1$, this expression reverts to $$\phi^n(p)(\omega)=\frac{\psi(\omega)^np(\omega)}{\sum_{\omega'}\psi(\omega')^np(\omega')},$$ from which we obtain our result in this case.

On the other hand, if $\alpha >1$,  all states $\omega$ for which $p(\omega)^{\alpha-1}\psi(\omega)$ are maximal get a positive probability in the limit, and all remaining states get zero probability.  For states for which $p(\omega)^{\alpha-1}\psi(\omega)$ is maximal, we have $\phi^*(p)(\omega)\propto_{\Omega} p(\omega)$.  But  since for any two states $\omega,\omega'$ for which $p(\omega)^{\alpha-1}\psi(\omega)$ is maximal, we have $p(\omega)^{\alpha-1}\psi(\omega)=p(\omega')^{\alpha-1}\psi(\omega')$, we have $\frac{p(\omega)}{p(\omega')}= \frac{\psi(\omega)^{\frac{1}{1-\alpha}}}{\psi(\omega')^{\frac{1}{1-\alpha}}}$.  Thus $\phi^*(p)(\omega)\propto_{\Omega} \psi(\omega)^{\frac{1}{1-\alpha}}$ for all states $\omega$ for which $\phi^*(\omega)>0$, and we are done. $\Box$

\end{document}